\documentclass{preprint}
\usepackage{geometry}
\usepackage{color}
\definecolor{note_fontcolor}{rgb}{1, 1, 1}
\definecolor{shadecolor}{rgb}{1, 1, 1}
\usepackage{verbatim}
\usepackage{calc}
\usepackage{framed}
\usepackage{mathrsfs}
\usepackage{amsthm}
\usepackage{amsmath}
\usepackage{amssymb}
\usepackage{esint}

\makeatletter

\numberwithin{equation}{section}
\theoremstyle{plain}
\newtheorem{thm}{\protect\theoremname}[section]
\newenvironment{lyxlist}[1]
{\begin{list}{}
{\settowidth{\labelwidth}{#1}
 \setlength{\leftmargin}{\labelwidth}
 \addtolength{\leftmargin}{\labelsep}
 }}
{\end{list}}
  \newtheorem{lem}[thm]{\protect\lemmaname}
  \newtheorem{prop}[thm]{\protect\propositionname}
  \newtheorem{cor}[thm]{\protect\corollaryname}
  \newtheorem{defn}[thm]{\protect\definitionname}

\usepackage{fancybox}
\usepackage{textcomp}
\usepackage[utf8]{inputenc}
\usepackage{lmodern}

\usepackage{stmaryrd}
\usepackage{hyperref}

\usepackage{color}
\usepackage{bbm}
\definecolor{magenta}{RGB}{30, 0, 50}
\definecolor{shadecolor}{rgb}{1, 1, 1}
\DeclareFontFamily{U}{mathx}{\hyphenchar\font45}
\DeclareFontShape{U}{mathx}{m}{n}{
      <5> <6> <7> <8> <9> <10>
      <10.95> <12> <14.4> <17.28> <20.74> <24.88>
      mathx10
      }{}
\DeclareSymbolFont{mathx}{U}{mathx}{m}{n}

\DeclareMathSymbol{\bigtimes}       {1}{mathx}{"91}%"



\makeatother

  \providecommand{\corollaryname}{Corollary}
  \providecommand{\definitionname}{Definition}
  \providecommand{\lemmaname}{Lemma}
  \providecommand{\propositionname}{Proposition}
\providecommand{\theoremname}{Theorem}

\begin{document}

\title{Glauber--Sudarshan-type quantizations and their path integral representations
for compact Lie groups}

\author{Hideyasu Yamashita}
\institute{Division of Liberal Arts and Sciences, Aichi-Gakuin University
 \\ \email{yamasita@dpc.aichi-gakuin.ac.jp}}
\maketitle

%% disable shaded env
%\renewenvironment{shaded}
%  {\bgroup\ignorespaces}
%  {\ignorespacesafterend\egroup} 

\newenvironment{trivenv}
  {\bgroup\ignorespaces}
  {\ignorespacesafterend\egroup}

\newcommand{\displabel}[1]{}

\newcommand{\hidable}[3]{#2}
\newcommand{\hidea}[1]{{#1}}
\newcommand{\hideb}[1]{{#1}}
\newcommand{\hidec}[1]{{#1}}
\newcommand{\hidep}[1]{{#1}}
\renewcommand{\hidec}[1]{}
\renewcommand{\hidep}[1]{}

\newcommand{\thlab}[1]{{\tt [#1]}}

\newenvironment{proofbar}
{\begin{leftbar}\noindent{\bf Proof.}}
{\noindent{\bf QED}\end{leftbar}}

\global\long\def\N{\mathbb{N}}
\global\long\def\C{\mathbb{C}}
\global\long\def\Z{\mathbb{Z}}
 \global\long\def\R{\mathbb{R}}
 \global\long\def\im{\mathrm{i}}

\global\long\def\di{\partial}
 \global\long\def\d{{\rm d}}

\global\long\def\ol#1{\overline{#1}}
\global\long\def\ul#1{\underline{#1}}
\global\long\def\ob#1{\overbrace{#1}}

\global\long\def\ov#1{\overline{#1}}

\global\long\def\then{\Rightarrow}
 \global\long\def\Then{\Longrightarrow}

\global\long\def\al{\alpha}
\global\long\def\de{\delta}
 \global\long\def\ep{\epsilon}
 \global\long\def\la{\lambda}
 \global\long\def\io{\iota}
 \global\long\def\th{\theta}
\global\long\def\si{\sigma}
 \global\long\def\om{\omega}

\global\long\def\De{\Delta}
 \global\long\def\Th{\Theta}
 \global\long\def\Om{\Omega}

\global\long\def\brho{\boldsymbol{\rho}}
\global\long\def\bDelta{\boldsymbol{\Delta}}
 \global\long\def\bmu{\boldsymbol{\mu}}
 \global\long\def\bchi{\boldsymbol{\chi}}
 \global\long\def\bPi{\boldsymbol{\Pi}}
 \global\long\def\bOm{\boldsymbol{\Omega}}

\global\long\def\cA{\mathcal{A}}
\global\long\def\cB{\mathcal{B}}
 \global\long\def\cC{\mathcal{C}}
 \global\long\def\cD{\mathcal{D}}
\global\long\def\cE{\mathcal{E}}
 \global\long\def\cF{\mathcal{F}}
 \global\long\def\cG{{\cal G}}
 \global\long\def\cH{\mathcal{H}}
 \global\long\def\cI{\mathcal{I}}
 \global\long\def\cJ{\mathcal{J}}
\global\long\def\cK{\mathcal{K}}
 \global\long\def\cL{\mathcal{L}}
 \global\long\def\cM{\mathcal{M}}
 \global\long\def\cN{\mathcal{N}}
 \global\long\def\cO{\mathcal{O}}
 \global\long\def\cP{\mathcal{P}}
 \global\long\def\cQ{\mathcal{Q}}
 \global\long\def\cR{\mathcal{R}}
 \global\long\def\cS{\mathcal{S}}
 \global\long\def\cT{\mathcal{T}}
 \global\long\def\cU{\mathcal{U}}
 \global\long\def\cV{\mathcal{V}}
 \global\long\def\cW{\mathcal{W}}
\global\long\def\cX{\mathcal{X}}
 \global\long\def\cY{\mathcal{Y}}
 \global\long\def\cZ{\mathcal{Z}}

\global\long\def\scA{\mathscr{A}}
\global\long\def\scB{\mathscr{B}}
\global\long\def\scC{\mathscr{C}}
\global\long\def\scD{\mathscr{D}}
 \global\long\def\scE{\mathscr{E}}
 \global\long\def\scF{\mathscr{F}}
 \global\long\def\scG{\mathscr{G}}
 \global\long\def\scH{\mathscr{H}}
 \global\long\def\scI{\mathscr{I}}
 \global\long\def\scJ{\mathscr{J}}
 \global\long\def\scK{\mathscr{K}}
 \global\long\def\scL{\mathscr{L}}
 \global\long\def\scM{\mathscr{M}}
 \global\long\def\scN{\mathscr{N}}
 \global\long\def\scO{\mathscr{O}}
 \global\long\def\scP{\mathscr{P}}
 \global\long\def\scR{\mathscr{R}}
\global\long\def\scS{\mathscr{S}}
 \global\long\def\scT{\mathscr{T}}
 \global\long\def\scU{\mathscr{U}}
 \global\long\def\scZ{\mathscr{Z}}

\global\long\def\bbA{\mathbb{A}}
 \global\long\def\bbB{\mathbb{B}}
 \global\long\def\bbD{\mathbb{D}}
 \global\long\def\bbF{\mathbb{F}}
 \global\long\def\bbG{\mathbb{G}}
 \global\long\def\bbI{\mathbb{I}}
 \global\long\def\bbK{\mathbb{K}}
 \global\long\def\bbL{\mathbb{L}}
 \global\long\def\bbM{\mathbb{M}}
 \global\long\def\bbP{\mathbb{P}}
 \global\long\def\bbQ{\mathbb{Q}}
 \global\long\def\bbT{\mathbb{T}}
 \global\long\def\bbU{\mathbb{U}}
 \global\long\def\bbX{\mathbb{X}}
 \global\long\def\bbY{\mathbb{Y}}
\global\long\def\bbW{\mathbb{W}}

\global\long\def\bbOne{1\kern-0.7ex  1}
 %defined as 1\kern-0.7ex1

\renewcommand{\bbOne}{\mathbbm{1}}

\global\long\def\bB{\mathbf{B}}
 \global\long\def\bG{\mathbf{G}}
 \global\long\def\bH{\mathbf{H}}
 \global\long\def\bM{\mathbf{M}}
 \global\long\def\bS{\boldsymbol{S}}
 \global\long\def\bT{\mathbf{T}}
 \global\long\def\bX{\mathbf{X}}
\global\long\def\bY{\mathbf{Y}}
\global\long\def\bW{\mathbf{W}}
 \global\long\def\boT{\boldsymbol{T}}

\global\long\def\fraka{\mathfrak{a}}
 \global\long\def\frakb{\mathfrak{b}}
 \global\long\def\frakc{\mathfrak{c}}
 \global\long\def\frake{\mathfrak{e}}
 \global\long\def\frakf{\mathfrak{f}}
 \global\long\def\fg{\mathfrak{g}}
 \global\long\def\frakh{\mathfrak{h}}
 \global\long\def\fraki{\mathfrak{i}}
 \global\long\def\frakk{\mathfrak{k}}
 \global\long\def\frakl{\mathfrak{l}}
 \global\long\def\frakm{\mathfrak{m}}
 \global\long\def\frakn{\mathfrak{n}}
 \global\long\def\frako{\mathfrak{o}}
 \global\long\def\frakp{\mathfrak{p}}
 \global\long\def\frakq{\mathfrak{q}}
 \global\long\def\frakr{\mathfrak{r}}
 \global\long\def\fs{\mathfrak{s}}
 \global\long\def\frakt{\mathfrak{t}}
 \global\long\def\fraku{\mathfrak{u}}

\global\long\def\fA{\mathfrak{A}}
 \global\long\def\fB{\mathfrak{B}}
 \global\long\def\fC{\mathfrak{C}}
 \global\long\def\fD{\mathfrak{D}}
 \global\long\def\fF{\mathfrak{F}}
 \global\long\def\fG{\mathfrak{G}}
 \global\long\def\fK{\mathfrak{K}}
 \global\long\def\fL{\mathfrak{L}}
 \global\long\def\fM{\mathfrak{M}}
 \global\long\def\fP{\mathfrak{P}}
 \global\long\def\fR{\mathfrak{R}}
 \global\long\def\fT{\mathfrak{T}}
 \global\long\def\fU{\mathfrak{U}}
 \global\long\def\fX{\mathfrak{X}}

\global\long\def\hM{\hat{M}}

\global\long\def\rM{\mathrm{M}}
\global\long\def\prj{\mathfrak{P}}

{} \global\long\def\sy#1{{\color{blue}#1}}

\global\long\def\magenta#1{{\color{magenta}#1}}

\global\long\def\symb#1{{\color{red}#1}}
{} %

\global\long\def\emhrb#1{\text{{\color{red}\huge{\bf #1}}}}

\newcommand{\symbi}[1]{\index{$ #1$}{\color{red}#1}} 

{} \global\long\def\SYM#1#2{\symb{#1}\index{${#1}$}_{\##2}}

\renewcommand{\SYM}[2]{\symb{#1}\index{${#1}$}}

\newcommand{\usuji}{\color[rgb]{0.7,0.4,0.4}} \newcommand{\usu}{\color[rgb]{0.5,0.2,0.1}}
\newenvironment{Usuji} {\begin{trivlist}   \item \usuji }  {\end{trivlist}}
\newenvironment{Usu} {\begin{trivlist}   \item \usu }  {\end{trivlist}} 

\newcommand{\term}[1]{\textcolor[rgb]{0, 0, 1}{\bf #1}}
\newcommand{\termi}[1]{\index{#1}\textcolor[rgb]{0, 0, 0.5}{\bf #1}}

\global\long\def\rG{\mathrm{G}}
 \global\long\def\rT{\mathrm{T}}
 \global\long\def\rH{\mathrm{H}}
 \global\long\def\rU{\mathrm{U}}

\global\long\def\supp{{\rm supp}}
\global\long\def\dom{\mathrm{dom}}
\global\long\def\ran{\mathrm{ran}}
 \global\long\def\leng{\text{{\rm leng}}}
 \global\long\def\diam{\text{{\rm diam}}}
 \global\long\def\Leb{\text{{\rm Leb}}}
 \global\long\def\meas{\text{{\rm meas}}}
\global\long\def\sgn{{\rm sgn}}
 \global\long\def\Tr{{\rm Tr}}
 \global\long\def\tr{\mathrm{tr}}
 \global\long\def\spec{{\rm spec}}
 \global\long\def\Ker{{\rm Ker}}
 \global\long\def\Lip{{\rm Lip}}
 \global\long\def\Id{{\rm Id}}
 \global\long\def\id{{\rm id}}

\global\long\def\ex{{\rm ex}}
 \global\long\def\Pow{\mathsf{P}}
 \global\long\def\Hom{\mathrm{Hom}}
 \global\long\def\grad{\mathrm{grad}}
 \global\long\def\End{{\rm End}}
 \global\long\def\Aut{{\rm Aut}}

\newcommand{\slim}{\mathop{\mbox{\rm s-lim}}} %

\newcommand{\wlim}{\mathop{\mbox{\rm w-lim}}}

\newcommand{\limsub}{\mathop{\mbox{\rm lim-sub}}}

\global\long\def\bboxplus{\boxplus}

\renewcommand{\bboxplus}{\mathop{\raisebox{-0.8ex}{\text{\begin{trivenv}\LARGE{}$\boxplus$\end{trivenv}}}}}

\global\long\def\shuff{\sqcup\kern-0.3ex  \sqcup}

\renewcommand{\shuff}{\shuffle}

\global\long\def\upha{\upharpoonright}

\global\long\def\ket#1{|#1\rangle}
 \global\long\def\bra#1{\langle#1|}

{} \global\long\def\lll{\vert\kern-0.25ex  \vert\kern-0.25ex  \vert}
 \renewcommand{\lll}{{\vert\kern-0.25ex  \vert\kern-0.25ex  \vert}}

\global\long\def\biglll{\big\vert\kern-0.25ex  \big\vert\kern-0.25ex  \big\vert\kern-0.25ex  }
 \global\long\def\Biglll{\Big\vert\kern-0.25ex  \Big\vert\kern-0.25ex  \Big\vert}

\newcommand{\iiia}[1]{{\left\vert\kern-0.25ex\left\vert\kern-0.25ex\left\vert #1
  \right\vert\kern-0.25ex\right\vert\kern-0.25ex\right\vert}}

\global\long\def\iii#1{\iiia{#1}}

\global\long\def\Upa{\Uparrow}
 \global\long\def\Nor{\Uparrow}

\newcommand{\vertt}{\kern-0.6ex\vert}
\renewcommand{\Nor}{[\kern-0.16ex ]}

\global\long\def\Prob{\mathbb{P}}
\global\long\def\Var{\mathrm{Var}}
\global\long\def\Cov{\mathrm{Cov}}
\global\long\def\Ex{\mathbb{E}}
{} %\newcommand{\F}{\mathbf{F}}
\global\long\def\Ae{{\rm a.e.}}
 \global\long\def\samples{\bOm}

\global\long\def\var{\textrm{{\rm var}}}
\global\long\def\Hol{\text{{\rm Höl}}}
 \global\long\def\hvar{\textrm{{\rm -var}}}
\global\long\def\hHol{\text{{\rm -Höl}}}

\global\long\def\pvar{p\textrm{{\rm -var}}}
\global\long\def\pHol{1/p\text{{\rm -Höl}}}
\global\long\def\frakt{\mathfrak{t}}

\global\long\def\var{\textrm{{\rm var}}}
\global\long\def\Hol{\text{{\rm Höl}}}
 \global\long\def\hvar{\textrm{{\rm -var}}}
\global\long\def\hHol{\text{{\rm -Höl}}}

\global\long\def\rpvar{\mathfrak{p}}
 \global\long\def\rpHol{\mathfrak{h}}

\global\long\def\prodlt{\prec}
 \global\long\def\prodgt{\succ}
 \global\long\def\prodr{\odot}

\global\long\def\prodltt{\prec\prec}

\renewcommand{\prodltt}{\prec\!\!\!\prec}

\global\long\def\bOne{{\bf 1}}

\global\long\def\Disk{\mathbb{D}^{2}}
\global\long\def\hcG{\hat{\mathcal{G}}}
\global\long\def\sfC{\mathsf{C}}

{} \global\long\def\crv{\mathfrak{c}}
\global\long\def\Crv{\mathfrak{C}}
 \global\long\def\gE{\mathrm{e}}
 \global\long\def\Rot{{\rm Rot}}

\global\long\def\bbm#1{\mathbbm{#1}}

\global\long\def\Mat{{\rm Mat}}

\global\long\def\cbo{{\bf c}}
 \global\long\def\reg{{\rm reg}}

\global\long\def\decoFor{\mathsf{DF}}
 \global\long\def\DF{\mathsf{DF}}

{} \global\long\def\modsp{\scT}
\global\long\def\regStr{\boldsymbol{T}}

\global\long\def\smoothfuncs{\scC}
 \global\long\def\jj{\mathtt{j}}
 \global\long\def\scriptj{\mathtt{j}}

\global\long\def\newNode{\circledast}
 \global\long\def\scriptf{\mathtt{f}}
 \global\long\def\scripth{\mathtt{h}}

\global\long\def\p{\mathbf{p}}
 \global\long\def\q{\mathbf{q}}
 \global\long\def\bA{\mathbf{A}}
 \global\long\def\x{\mathbf{x}}

\global\long\def\div{\mathrm{div}}
 \global\long\def\be{\beta}
 \global\long\def\La{\Lambda}
 \global\long\def\Ga{\Gamma}

\global\long\def\wick#1{:\!#1\!:}
 \global\long\def\dag{\dagger}

\global\long\def\braket#1{\langle#1\rangle}
 \global\long\def\ka{\kappa}
 \global\long\def\z{\mathbf{z}}

\global\long\def\tI{t_{\mathrm{I}}}
 \global\long\def\tF{t_{\mathrm{F}}}

\global\long\def\ActionIntegral{\mathrm{AI}}

\global\long\def\Hc{h}

\global\long\def\coherents{\mathbf{c}}
 \global\long\def\ssW{\mathsf{W}}

\global\long\def\Ten{\bullet}
{} %

\global\long\def\TT{\intercal}
 \renewcommand{\TT}{\mathsf{T}}

\global\long\def\trit{\vartriangle\!\! t}

\global\long\def\Killing{{\rm \boldsymbol{\kappa}}}
 \global\long\def\spec{{\rm spec}}

\global\long\def\nnn{\mathfrak{k}}

\global\long\def\Ad{{\rm Ad}}

\global\long\def\Gxz{G\cdot x_{0}}
 \global\long\def\lbundle{\scL_{\lambda}}
 \global\long\def\Hilb{\cH_{\lambda}}
 \global\long\def\Image{{\rm Im}}
 \global\long\def\tautlog{{\rm taut}}
 \global\long\def\sphere{\mathbb{S}}
 \global\long\def\Proj{\mathbf{pj}}
 \global\long\def\telem{\mathbf{t}}
 \global\long\def\sectio{\mathbf{sect}}

\global\long\def\hwvec{\mathbf{v}_{\lambda}}
{} \global\long\def\Hwvproj{{\bf E}_{\lambda}}

\global\long\def\lwvec{{\bf w}_{\lambda}}

\global\long\def\lbundle{\mathscr{L}_{\lambda}}
 \global\long\def\Gxz{G\cdot x_{0}}

\global\long\def\Rtrans{{\rm Rt}}
 \global\long\def\Ltrans{{\rm Lt}}

\global\long\def\Rght{{\rm R}}
 \global\long\def\Lft{{\rm L}}
 \global\long\def\Casi{{\bf c}}
 \global\long\def\Rtrans{\mathscr{T_{\Rght}}}
 \global\long\def\Ltrans{\mathscr{T_{{\rm \Lft}}}}
 \global\long\def\Maurer{\mathscr{M}}
 \global\long\def\du{\underline{{\rm d}}}
 \global\long\def\bphi{\boldsymbol{\varphi}}

{} \global\long\def\ss{{\bf r}}
 \global\long\def\infspec{{\bf c}_{\lambda}}

\global\long\def\Roots{\boldsymbol{R}}
 \global\long\def\Sn{{\rm sig}}
 \global\long\def\lift{{\rm lift}}
 \global\long\def\bpi{\boldsymbol{\pi}}

\global\long\def\Tensor{\boldsymbol{T}}
 \global\long\def\bOmega{\boldsymbol{\Omega}}
 \global\long\def\VECSP{\mathbb{V}}
 \global\long\def\projection{{\rm pr}}
 \global\long\def\dissect{\cD}
 \global\long\def\bUpsilin{\boldsymbol{\Upsilon}}
 \global\long\def\GEOMR{\mathbf{GR}}

\global\long\def\fextend#1{\hat{#1}}
 \global\long\def\eigenval{\varepsilon}
\global\long\def\symb#1{{\color[rgb]{0, 0, 0}#1}}
\global\long\def\term#1{\textcolor[rgb]{0, 0, 0.2}{\bf #1}}
\global\long\def\SYM#1#2{\symb{#1}\index{${#1}$}_{\##2}}

% disable shaded env
\renewenvironment{shaded}
  {\bgroup\ignorespaces}
  {\ignorespacesafterend\egroup}

\begin{abstract}

In this paper, we consider an arbitrary irreducible unitary representation
$(\pi_{\lambda},V_{\lambda})$ of a compact connected, simply connected
semisimple Lie group $\rG$ with highest weight $\lambda$, and apply
the idea of Daubechies--Klauder (1985) and Yamashita (2011) on rigorous
coherent-state path integrals to this representation, where the orbit
of the highest weight vector is interpreted as the manifold of coherent
states. Our main theorem is two-fold: the first main theorem is in
terms of Brownian motions and stochastic integrals, and proven using
the Feynman--Kac--It\^o formula on a vector bundle of a Riemannian
manifold, due to G\"uneysu (2010). In the second main theorem, we
consider a sequence $(\mu_{n})$ of finite measures on the space of
smooth paths, and a `path integral' is defined to be a limit of the
integrals with respect to $(\mu_{n})$. The formulation and the proof
of the second main theorem employ \emph{rough path theory} originated
by Lyons (1998).
\end{abstract}

\section{Introduction}

There are several approaches to mathematical foundation of path integrals
occurring in quantum physics. Feynman's original idea \cite{FH65}
is to represent the time evolution of a quantum system, as well as
the expectation values of observables in it, by an integral on the
space of paths on the\emph{ configuration space} of the system. As
is well known, if we consider the ``imaginary time'' evolution instead
of real time evolution, so-called the \emph{Wick rotation}, a large
part of the idea can be made rigorous by the Feynman--Kac theorem
and its generalizations, and this ``imaginary time $+$ Feynman--Kac''
approach is the most successful one. However, note that in the imaginary-time
approaches, it is difficult to deal with time-dependent Hamiltonians,
as well as non-unitary time evolutions occurring in open systems.
This implies that it is hard to apply the imaginary-time methods to
e.g. the theories of quantum information/probability, where time-dependent
Hamiltonians and non-unitary time evolutions (e.g. decoherences) frequently
occur.

On the other hand, the notion on configuration-space path integrals
are believed to be derived from more general notion of \emph{phase-space}
path integrals. Although configuration-space path integrals are preferred
to phase-space path integrals especially in relativistic quantum field
theories for their `manifest Lorentz covariance,' the latter ones
will be more fundamental if we consider a path integral as a procedure
of quantization of a classical system; The main stream of the rigorous
studies of quantization (e.g. the theories of geometric/deformation
quantization) are formulated on phase spaces. Unlike imaginary-time
configuration-space path integrals, little is known about the rigorous
justification of general phase-space path integrals (in real or imaginary
time).

There is another notion of \emph{coherent-state} path integrals, which
resembles to that of phase-space path integrals; Sometimes the former
notion is said to be a \emph{part} of the latter one, but the precise
relation between them is not clear since the rigorous definitions
of both have not been given. The notion of \emph{coherent states}
are introduced by Glauber \cite{Gla63}, and later generalized by
many authors. The original `usual' coherent states are called \emph{Glauber
coherent states} (GCS), to distinguish them from others. Although
no widespread rigorous definition of generalized coherent states seems
to exist, it is commonly recognized that if a unitary highest weight
irreducible representation of a transformation group of a system is
given, the orbit of the highest weight vector is a typical example
of the manifold of coherent states (see e.g. \cite{Lis91}). 

In 1985, Daubechies and Klauder \cite{DK85} gave a rigorous GCS path
integral formula representing real-time evolution for some class of
Hamiltonians, in terms of Brownian motions and stochastic integrals.
Yamashita \cite{Yam11} studied GCS path integrals in a similar idea
but for other class of Hamiltonians, and with an emphasis on geometric
meaning of them. Although an imaginary-time configuration-space path
integral can be defined as an integral with respect to a single Wiener
measure by the Feynman--Kac theorem, it appears that a path integral
of other kinds cannot be defined to be an integral with respect to
a single Borel measure. Instead we consider a sequence $(\mu_{n})_{n\in\N}$
of measures, and regard a path integral as a limit of the form
\[
\lim_{n\to\infty}\int F(\psi)\d\mu_{n}(\psi).
\]

In this paper, we consider an arbitrary irreducible unitary representation
of a compact connected, simply connected semisimple Lie group $\rG$,
and apply the idea of \cite{Yam11} to the orbit of the highest weight
state $\rG\cdot\Hwvproj$, which is a symplectic manifold with the
natural symplectic 2-form $\omega$, called the \emph{Kirillov--Kostant--Souriau
2-form}, identifying the orbit $\rG\cdot\Hwvproj$ with the coadjoint
orbit $\rG\cdot\lambda$. Thus $\left(\rG\cdot\Hwvproj,\omega\right)$
can be regarded as a phase space of some classical-mechanical system.
However, here we shall deal with the integral on the space of paths
on $\rG$, not on $\rG\cdot\Hwvproj$. The main reason for that is
as follows. Consider the usual flat phase space $M=\R^{2n}$ with
a symplectic 2-form $\omega$. Then there exists a 1-form $\theta$,
called the \emph{canonical 1-form}, such that $\d\theta=\omega$.
If a path $C$ on $M$ is given, we can consider the line integral
$\int_{C}\theta$, interpreted as the ``action along $C$.'' On
the other hand, for general symplectic manifold $(M,\omega)$, the
1-form $\theta$ satisfying $\d\theta=\omega$ may not exist; Even
if such $\theta$ exists, the reason for choosing a distinguished
$\theta$, which should be called a `natural' or `canonical' one,
may not exist. However, a `fairly natural' 1-form $\theta$ exists
on $\rG$, not on $\rG\cdot\lambda\cong\rG\cdot\Hwvproj$; that is,
$\theta$ is the left-invariant 1-form (i.e. the \emph{Maurer--Cartan}
form) w.r.t. the highest weight $\lambda$. Let $\tilde{\omega}$
be the pullback of $\omega$ w.r.t the map $\rG\ni g\mapsto g\cdot\lambda\in\rG\cdot\lambda$,
then we find $\tilde{\omega}=-d\theta$. Thus our path integral can
be said to be nearly a coherent-state or phase-space path integral,
but not exactly.

The paper is organized as follows. In Section \ref{sec:mainTheorem},
the statement of the main theorem, together with the definitions of
notions (including the GS quantization) and symbols needed to state
them, is presented,. Our main theorem is two-fold: the first theorem
\ref{thm:Main-Brownian} is in terms of Brownian motions and stochastic
integrals, and the second theorem \ref{thm:Main-smooth} is formulated
as a limit of the integrals on the space of smooth paths. In Section
\ref{sec:pre-Borel-Weil}, we define the subspace $\cH_{\lambda}(\rG)\subset L^{2}(\rG)$,
and state the ``pre-Borel--Weil theorem'' on $\cH_{\lambda}(\rG)$,
essentially used in the proof of the main theorem. The \emph{Borel--Weil
theorem,} which is a complex-geometric representation of the irreducible
unitary representations of $\rG$, is derived from the pre-Borel--Weil
theorem, but we need only the latter theorem in this paper. In Section
\ref{sec:Casimir} and \ref{sec:magnetic}, we define the magnetic
Laplacian $\Delta^{\alpha}$ on $\rG$, and represent $\cH_{\lambda}(\rG)$
as a ``ground state space'' of $\Delta^{\alpha}$. In Section \ref{sec:GSquantOnH},
we prove the theorem which state that any GS quantization is represented
as a projection onto $\cH_{\lambda}(\rG)$. In Section \ref{sec:asymp},
we prove the asymptotic representation of the (real-)time evolution
of GS-quantized system, in terms of $\Delta^{\alpha}$. In Section
\ref{sec:pathInt-Brown}, we prove the first main theorem. In Section
\ref{sec:roughPath} and \ref{sec:BrownAsRP}, we present an outline
of \emph{rough path theory} in the style of \cite{FV10b}. In Section
\ref{sec:pathInt-smooth}, we prove the second main theorem.

\section{Main theorem}

\label{sec:mainTheorem}\displabel{sec:mainTheorem}

First we recall basic definitions on Lie groups and Lie algebras which
we will use in this paper. 

Let $\rG$ be a compact connected, simply connected semisimple Lie
group, that is, $\rG$ be one of ${\rm SU}(n)\ (n\ge2)$, ${\rm Spin}(n)\ (n\ge3)$,
${\rm Sp}(n)\ (n\ge1)$ and the five exceptional groups of the types
$E_{6}$, $E_{7}$, $E_{8}$, $F_{4}$ and $G_{2}$. Let $\fg$ be
the Lie algebra of $\rG$; $\rG_{\C}$ and $\fg_{\C}$ be the complexifications
of $\rG$ and $\fg$, respectively. Fix a maximal torus $\rT\subset\rG$
(i.e. $\rT$ is a maximal commutative connected compact subgroup of
$\rG$. In fact $\rT\cong\rU(1)^{\ell}$ for some $\ell$). The Lie
algebra of $\rT$ is denoted by $\frakt$, and its complexification
by $\frakt_{\C}$ (the Cartan subalgebra of $\fg_{\C}$). Let $\ell$
be the \term{rank}\index{zzz@rank}\hypertarget{rank}{} of $\rG$, i.e. $\symb{\ell}\index{${l}$@${\ell}$}\hypertarget{ref}{}:=\dim\frakt$. Let
$\symb{\hat{\rG}}\index{${Ghat}$@${\hat{\rG}}$}\hypertarget{ref.1}{}$ denote the unitary dual of $\rG$, i.e. the
set of (the equivalence classes of ) the irreducible unitary presentations
of $\rG$. 

Let $\Killing(\bullet,\bullet)$ denote the Killing form on $\fg_{\C}$.
Define the linear bijection $\symb{\nu}\index{${nu}$@${\nu}$}\hypertarget{ref.2}{}:\fg_{\C}\to\fg_{\C}^{*}$
by $\nu(X)(Y):=\Killing(X,Y)$. Define the bilinear form $(\bullet,\bullet)$
on $\fg_{\C}^{*}$ by

\begin{equation}
\symb{(\alpha,\beta)}\index{${(\alpha,\beta)}$@${(\alpha,\beta)}$}\hypertarget{ref.3}{}:=\Killing(\nu^{-1}(\alpha),\nu^{-1}(\beta)),\qquad\alpha,\beta\in\fg_{\C}^{*}.\label{eq:def:(al,be)}\displabel{eq:def:(al,be)}
\end{equation}
For $\alpha\in\frakt_{\C}^{*}$, let 
\[
\symb{\fg_{\C}^{\alpha}}\index{${galp}$@${\fg_{\C}^{\alpha}}$}\hypertarget{ref.4}{}:=\left\{ X\in\fg_{\C}|\ [T,X]=\alpha(T)X,\quad\forall T\in\frakt_{\C}\right\} 
\]
and $\symb{\Roots}\index{${R}$@${\Roots}$}\hypertarget{ref.5}{}:=\left\{ \alpha\in\frakt_{\C}^{*}|\ \fg_{\C}^{\alpha}\neq\{0\}\right\} \setminus\{0\}$,
the set of \term{roots}\index{zzz@roots}\hypertarget{roots}{} of $\fg_{\C}$. Fix a decomposition $\Roots=\symb{\Roots^{+}}\index{${R+}$@${\Roots^{+}}$}\hypertarget{ref.6}{}\cup\symb{\Roots^{-}}\index{${R-}$@${\Roots^{-}}$}\hypertarget{ref.7}{}$,
$\Roots^{+}\cap\Roots^{-}=\emptyset$ such that $\alpha\in\Roots^{+}$
iff $-\alpha\in\Roots^{-},$ and that
\[
\alpha,\beta\in\Roots^{+},\ \alpha+\beta\in\Roots\Longrightarrow\alpha+\beta\in\Roots^{+}
\]
Each element of $\Roots^{+}$ is called a \term{positive root}\index{zzz@positive root}\hypertarget{positive+root}{}.
The subset $\Roots_{{\rm s}}^{+}\subset\Roots^{+}$ of \term{simple roots}\index{zzz@simple roots}\hypertarget{simple+roots}{}
is defined by

\[
\symb{\Roots_{{\rm s}}^{+}}\index{${R+s}$@${\Roots_{{\rm s}}^{+}}$}\hypertarget{ref.8}{}=\left\{ \alpha_{1},...,\alpha_{\ell}\right\} :=\left\{ \alpha\in\Roots^{+}|\ \alpha\neq\beta+\gamma,\ \forall\beta,\gamma\in\Roots^{+}\right\} \ 
\]
Define the \term{weight lattice}\index{zzz@weight lattice}\hypertarget{weight+lattice}{} by
\begin{align*}
\symb{P}\index{${P}$@${P}$}\hypertarget{ref.9}{} & :=\left\{ \la\in\im\frakt^{*}|\left(\al^{\vee},\la\right)\in\Z\quad\forall\al\in\Roots\right\} \subset i\frakt^{*}
\end{align*}
where $\symb{\alpha^{\vee}}\index{${alphavee}$@${\alpha^{\vee}}$}\hypertarget{ref.10}{}:=2\alpha/(\alpha,\alpha)$ is
the \term{coroot}\index{zzz@coroot}\hypertarget{coroot}{} corresponding to $\alpha.$ Each element of $P$
is called an \term{algebraically integral weight}\index{zzz@algebraically integral weight}\hypertarget{algebraically+integral+weight}{}.

Let $\ker\exp_{\frakt}:=\left\{ X\in\frakt|\ \exp(X)=1_{\rG}\right\} $
where $1_{\rG}$ is the unit in $\rG$. The \term{character lattice}\index{zzz@character lattice}\hypertarget{character+lattice}{}
for $\rT$ is defined by
\[
\symb{\cX(\rT)}\index{${X(T)}$@${\cX(\rT)}$}\hypertarget{ref.11}{}:=\{\la\in\im\frakt^{*}|\left\langle \la,X\right\rangle \in2\pi\im\Z\quad\forall X\in\ker\exp_{\frakt}\}.
\]
Each element of $\cX(\rT)$ is called an \term{analytically integral weight}\index{zzz@analytically integral weight}\hypertarget{analytically+integral+weight}{}.
Under the assumption that $\rG$ is simply connected, the character
lattice $\cX(\rT)$ equals the weight lattice $P$ (we have $\cX(\rT)\subset P$
in general). The set of \term{dominant weights}\index{zzz@dominant weights}\hypertarget{dominant+weights}{} $\cX_{+}(\rT)\subset\cX(\rT)$
is defined by
\[
\symb{\cX_{+}(\rT)}\index{${X+(T)}$@${\cX_{+}(\rT)}$}\hypertarget{ref.12}{}:=\left\{ \lambda\in\cX(\rT)|\ \left(\la,\al_{i}^{\vee}\right)\in\Z_{+},\quad i=1,...,\ell\right\} 
\]
It is shown that there is a one-to-one correspondence between $\hat{\rG}$
and $\cX_{+}(\rT)$; For each $\la\in\cX_{+}(\rT)$, let $(\symb{\d\pi_{\lambda}}\index{${dpi}$@${\d\pi_{\lambda}}$}\hypertarget{ref.13}{},V_{\lambda})$
be the irreducible highest weight representation of $\fg_{\C}$ on
the complex vector space $V_{\lambda}$ with highest weight $\la$.
Let $\symb{\pi_{\lambda}}\index{${pilam}$@${\pi_{\lambda}}$}\hypertarget{ref.14}{}$ denote the lift of $\d\pi_{\lambda}$
to an irreducible representation of $\rG$, i.e. $(\pi_{\lambda},V_{\lambda})$
be the irreducible representation of $\rG$ such that $\d\pi_{\lambda}$
is the differential representation of $\pi_{\lambda}$. Each $V_{\lambda}$
has the inner product $\left\langle \bullet|\bullet\right\rangle $
where $\pi_{\lambda}$ is unitary. (We use the notation $\left\langle \bullet|\bullet\right\rangle $
only for usual (positive-definite) inner products, linear in the second
variable; on the other hand the notation $\left\langle \bullet,\bullet\right\rangle $
denotes more generic forms, possibly not positive-definite.) We often
write the representation $\d\pi_{\lambda}$ of $\fg_{\C}$ simply
as $\pi_{\lambda}$, unless confusion arises.

Let $\hwvec\in V_{\lambda}$ ($\left\Vert \hwvec\right\Vert =1$)
be a highest weight vector, i.e.%
\[
\pi_{\lambda}(X)\hwvec=\lambda(X)\hwvec,\qquad\forall X\in\frakt.
\]
For $v\in V_{\lambda}$, define $v^{*}\in V_{\lambda}^{*}$ by $v^{*}(u):=\left\langle v|u\right\rangle $,
$u\in V_{\lambda}$. Let $\symb{\Hwvproj}\index{${Elamb}$@${\Hwvproj}$}\hypertarget{ref.15}{}=\hwvec\hwvec^{*}$
be the orthogonal projection from $V_{\lambda}$ onto $\C\hwvec$.
Let
\[
\symb{g\cdot\Hwvproj}\index{${g\cdot\Hwvproj}$@${g\cdot\Hwvproj}$}\hypertarget{ref.16}{}:=\pi_{\lambda}(g)\Hwvproj\pi_{\lambda}(g^{-1}),\quad g\in\rG,
\]
and $\symb{\rG\cdot\Hwvproj}\index{${G.E}$@${\rG\cdot\Hwvproj}$}\hypertarget{ref.17}{}:=\left\{ g\cdot\Hwvproj:\ g\in\rG\right\} $,
called the \term{orbit}\index{zzz@orbit}\hypertarget{orbit}{} through $\Hwvproj$, or the manifold of
\term{coherent states}\index{zzz@coherent states}\hypertarget{coherent+states}{} in the physical context (see e.g. \cite{Lis91}). 

For $h\in C^{\infty}(\rG\cdot\Hwvproj,\R)$, define $\fextend h\in C^{\infty}(\rG,\R)$
by $\symb{\fextend h}\index{${hhat}$@${\fextend h}$}\hypertarget{ref.18}{}(g):=h(g\cdot\Hwvproj)$, and define the
operator $\cQ(h)$ by

\[
\symb{\cQ(h)}\index{${Q(f)}$@${\cQ(h)}$}\hypertarget{ref.19}{}:=d_{\lambda}\int_{G}\fextend h(g)\left(g\cdot\Hwvproj\right)\d g,\qquad\symb{d_{\lambda}}\index{${dlam}$@${d_{\lambda}}$}\hypertarget{ref.20}{}:=\dim V_{\lambda}
\]
where $\d g$ denotes the Haar measure on $\rG$, normalized so that
$\int_{\rG}\d g=1$. We call the map $\cQ:C^{\infty}(\rG\cdot\Hwvproj)\to\End(V_{\lambda})$
the \term{Glauber--Sudarshan-type quantization}\index{zzz@Glauber--Sudarshan-type quantization}\hypertarget{Glauber--Sudarshan-type+quantization}{} (or simply, the
\term{GS quantization}\index{zzz@GS quantization}\hypertarget{GS+quantization}{}). If $h$ is a real-valued , then the GS
quantization $\cQ(h)$ is self-adjoint, and so $\left\{ e^{\im t\cQ(h)}|t\in\R\right\} $
is a one-parameter unitary group. Note that every self-adjoint operator
on $V_{\lambda}$, possibly not in $\pi_{\lambda}(\im\fg)$, is represented
as $\cQ(h)$ for some $h\in C^{\infty}(\rG\cdot\Hwvproj,\R)$. (This
naming is by an analogue of the \emph{Glauber--Sudarshan representation}
(also called the $P$-\emph{representation}) for the Glauber coherent
states, frequently used in quantum optics. A mathematical reason for
calling $\cQ(h)$ a ``\emph{quantization} of $h$'' is seen in e.g.
\cite{Lis91,Lan98}.) 

For each $v\in V_{\lambda}$, define $\tilde{v}\in L^{2}(\rG)$ by
\[
\symb{\tilde{v}}\index{${vhat}$@${\tilde{v}}$}\hypertarget{ref.21}{}(g):=d_{\lambda}^{1/2}\left\langle \pi_{\lambda}(g)\hwvec|v\right\rangle ,\qquad g\in\rG.
\]
Then the map $v\mapsto\tilde{v}$ turns out to be an isometry.

Since the Killing form $\Killing$ is negative-definite on $\fg$,
$\left\langle X|Y\right\rangle _{\fg}:=-\Killing(X,Y)$ defines an
inner product on $\fg$. This induces a Riemannian metric on $\rG$.
Now consider the Brownian motion $B$ on the Riemannian manifold $\rG$
in the time interval $[0,\infty)$, where the distribution of the
starting point is uniform on $\rG$, i.e. equals the Haar measure
$\d g$ on $\rG$. Let $\symb{\mu^{1}}\index{${mu1}$@${\mu^{1}}$}\hypertarget{ref.22}{}$ be a probability measure
on $C([0,\infty),\rG)$ which is the law of this Brownian motion (i.e.
a Wiener measure uniform on $\rG$). 

{} For $\ss>0$, define the probability measure $\mu^{\ss}$ on $C([0,\infty),\rG)$
by
\[
\d\symb{\mu^{\ss}}\index{${muss}$@${\mu^{\ss}}$}\hypertarget{ref.23}{}(B):=\d\mu(B(\ss^{-1}\bullet)),
\]
i.e., $\mu^{\ss}$ is the time rescaling of $\mu^{1}$, so that the
$\mu^{\ss}$-Brownian motion diffuses $\ss$ times faster than the
$\mu^{1}$-Brownian motion.

\begin{flushleft}
For $\alpha\in\fg_{\C}^{*}$ define the $\C$-valued 1-form $\symb{\alpha^{\Rght}}\index{${alphaR}$@${\alpha^{\Rght}}$}\hypertarget{ref.24}{}$
on $\rG$ as the unique right-invariant $\C$-valued 1-form such that
$\alpha_{1_{\rG}}^{\Rght}=\alpha|_{\fg}$. If $\rG$ is embedded in
the matrix Lie group ${\rm GL}(n,\C)$, we have
\begin{equation}
\alpha_{g}^{\Rght}(X)=\alpha\left(X_{g}g^{-1}\right),\qquad g\in\rG,\ X\in\mathfrak{X}(\rG)\label{eq:def:alphaR}\displabel{eq:def:alphaR}
\end{equation}
where $\mathfrak{X}(\rG)$ is the space of vector fields on $\rG$.
We naturally view $\frakt^{*}$ as a subspace of $\fg^{*}$, and $\fg^{*}$
as a real linear subspace of $\fg_{\C}^{*}$ by
\[
\fg^{*}\ni\alpha\mapsto\alpha'\in\fg_{\C}^{*},\qquad\alpha'(X+\im Y):=\alpha(X)+\im\alpha(Y),\quad X,Y\in\fg.\ \ 
\]
Hence we have $\im\frakt\hookrightarrow\fg_{\C}^{*}$, and so $\alpha^{\Rght}$
is defined for any $\alpha\in\im\frakt^{*}$, which is a $\im\R$-valued
1-form.
\par\end{flushleft}

Let $\rho\in\im\frakt^{*}$ be the half sum of positive roots of $\fg_{\C}$:
$\symb{\rho}\index{${rho}$@${\rho}$}\hypertarget{ref.25}{}:=\frac{1}{2}\sum_{\alpha\in\Roots_{+}}\alpha.$ %
For $h\in C^{\infty}(\rG\cdot\Hwvproj,\R)$ and $t\ge0$, define the
$\C$-valued random variable $\cI_{t}(h)\equiv\cI_{t}(h;B)$ by 
\[
\symb{\cI_{t}(h;B)}\index{${Ih0t}$@${\cI_{t}(h;B)}$}\hypertarget{ref.26}{}:=\int_{0}^{t}\alpha^{\Rght}(\du B_{s})-\im\int_{0}^{t}\fextend h(B_{s})\d s,\quad\symb{\alpha}\index{${alph}$@${\alpha}$}\hypertarget{ref.27}{}:=-(\lambda+\rho),
\]
where the line integral $\int_{0}^{t}\alpha^{\Rght}(\du B_{s})$ is
a stochastic integral in the sense of Stratonovich, and $\symb{\fextend h}\index{${ftilde}$@${\fextend h}$}\hypertarget{ref.28}{}(g):=h(g\cdot\Hwvproj)$
for $g\in\rG$. Note that $\cI_{t}(h)\in\im\R$, and so $e^{\cI_{t}(h)}\in{\rm U}(1)$.
Fix an arbitrary $v_{1}\in V_{\lambda}$ with $\left\Vert v_{1}\right\Vert =1$,
and set

\[
\symb{Z_{\lambda,t,\ss}}\index{${Zlambt}$@${Z_{\lambda,t,\ss}}$}\hypertarget{ref.29}{}:=\int_{C([0,\infty),\rG)}\left[e^{\cI_{t}(0;B)}\ol{\tilde{v}_{1}(B_{0})}\tilde{v}_{1}(B_{t})\right]\d\mu^{\ss}(B).
\]
It is shown that $Z_{\lambda,t,\ss}>0$, and that $Z_{\lambda,t,\ss}$
does not depend on $v_{1}$.

\begin{shaded}%
\begin{thm}
[Main: Brownian form] \label{thm:Main-Brownian}\displabel{thm:Main{-}Brownian}Let $h\in C^{\infty}(\rG\cdot\Hwvproj,\R)$
be a `classical Hamiltonian.' Then for any $u,v\in V_{\lambda}$ and
$t>0,$ we have
\begin{equation}
\bigl\langle u|e^{\im t\cQ(h)}v\bigr\rangle=\lim_{\ss\to\infty}\int_{C([0,\infty),\rG)}\left[e^{\cI_{t}(h;B)}\ol{\tilde{u}(B_{0})}\tilde{v}(B_{t})\right]\frac{\d\mu^{\ss}(B)}{Z_{\lambda,t,\ss}}.\ \label{eq:main-brown}\displabel{eq:main{-}brown}
\end{equation}
\end{thm}
\end{shaded}%

Consider the problem of generalizing this result to the cases where 
\begin{lyxlist}{(ii)}
\item [{(i)}] $\rG$ is a finite-dimensional non-compact Lie group;
\item [{(ii)}] $\rG$ is an infinite-dimensional non-compact Lie group
(e.g. infinite-dimensional Heisenberg group, spin group, gauge transformation
group, etc.)
\end{lyxlist}
In both case, the representation space $V_{\lambda}$ is infinite-dimensional.

In case (i), if $\rG$ has an invariant Riemannian metric ${\rm g}$,
the `standard' Brownian motion on $(\rG,{\rm g})$ exists, and so
it is conjectured that some equation similar to (\ref{eq:main-brown})
holds for an irreducible unitary representation of $\rG$. (Some positive
results concerning this conjecture are given in \cite{DK85,Yam11}
when $\rG$ is a finite-dimensional Heisenberg group.%
)

However, in the other cases of (i), and in all cases of (ii), the
\emph{standard} Brownian motion on $\rG$ does not exist. Hence any
straightforward generalization of (\ref{eq:main-brown}) seems impossible
in these cases. To make matters worse, $\rG$ have no invariant measure
in case (ii), and hence the left/right regular representations of
$\rG$ on $L^{2}(\rG)$ cannot be defined. Thus it is worth reformulating
Theorem \ref{thm:Main-Brownian} to a statement which refers to neither
Brownian motions nor $L^{2}(\rG)$:

\begin{shaded}%
\begin{thm}
[Main: smooth form]\label{thm:Main-smooth}\displabel{thm:Main{-}smooth} In the setting of Theorem
\ref{thm:Main-Brownian}, let $(\mu_{k})_{k\in\N}$ be a sequence
of finite measures on the smooth path space $C^{\infty}([0,\infty),\rG)$.
If $(\mu_{k})_{k\in\N}$ satisfies some conditions given in Sec.\ref{sec:pathInt-smooth},
then for each $u,v\in V_{\lambda}$ and $t>0$, 
\begin{equation}
\bigl\langle u|e^{\im t\cQ(h)}v\bigr\rangle=\lim_{k\to\infty}\int_{C^{\infty}([0,\infty),\rG)}\left[e^{\cI_{t}(h;\varphi)}\left\langle u|\pi_{\lambda}(\varphi(0))\hwvec\right\rangle \left\langle \pi_{\lambda}(\varphi(t))\hwvec|v\right\rangle \right]\d\mu_{k}(\varphi),\label{eq:main-smooth1}\displabel{eq:main{-}smooth1}
\end{equation}
where
\[
\symb{\cI_{t}(h;\varphi)}\index{${Ithphi}$@${\cI_{t}(h;\varphi)}$}\hypertarget{ref.30}{}:=\int_{0}^{t}\alpha^{\Rght}(\d\varphi(s))-\im\int_{0}^{t}\fextend h(\varphi(s))\d s,\quad\symb{\alpha}\index{${alph}$@${\alpha}$}\hypertarget{ref.31}{}:=-(\lambda+\rho).
\]
\end{thm}
\end{shaded}

Here we raise the problem to give a necessary and sufficient condition
for $\{\mu_{k}\}_{k\in\N}$ to satisfy Eq.(\ref{eq:main-smooth1}).
Although it seems quite difficult to give a perfect answer to this
problem, a fairly good sufficient condition is given in terms of \emph{rough
path theory,} originated by Lyons \cite{Lyo98}.

\section{Pre-Borel--Weil theorem}

\label{sec:pre-Borel-Weil}\displabel{sec:pre{-}Borel{-}Weil}

{} 

Let $\rG$ be a compact connected, simply connected semisimple Lie
group, and $\fg$ be the Lie algebra of $\rG$; $\rG_{\C}$ and $\fg_{\C}$
be the complexifications of $\rG$ and $\fg$, respectively. Fix a
maximal torus $\rT\subset\rG$, and its Lie algebra $\frakt$. 

Define the \term{adjoint operation}\index{zzz@adjoint operation}\hypertarget{adjoint+operation}{} on $\fg_{\C}$ to be the antilinear
map $*:\fg_{\C}\to\fg_{\C}$ such that $X^{*}=-X$ for all $X\in\fg$.
We see the relation $[X,Y]^{*}=[Y^{*},X^{*}]$, $X,Y\in\fg_{\C}$.
If $\fg_{\C}$ is embedded in the matrix Lie algebra $\Mat(n,\C)\cong\mathfrak{gl}(n,\C)$,
$X^{*}$ is nothing but the adjoint matrix of $X\in\fg_{\C}$.

Let

\[
\symb{\frakn^{-}}\index{${u-}$@${\frakn^{-}}$}\hypertarget{ref.32}{}:=\bigoplus_{\alpha\in\Roots^{+}}\fg_{\C}^{-\alpha},\qquad\symb{\frakb^{-}}\index{${b-}$@${\frakb^{-}}$}\hypertarget{ref.33}{}:=\frakt\oplus\frakn^{-}.
\]
Let $\symb{T_{1},...,T_{\ell}}\index{${Ti}$@${T_{1},...,T_{\ell}}$}\hypertarget{ref.34}{}\in\im\frakt$ %
be a basis of $\frakt_{\C}$ such that 
\[
\Killing(T_{i},T_{j})=\delta_{ij}.
\]
For each $\alpha\in\Roots$, we can take an element $\symb{E_{\alpha}}\index{${Ealp}$@${E_{\alpha}}$}\hypertarget{ref.35}{}\in\fg_{\C}^{\alpha}$
such that $E_{-\alpha}=-E_{\alpha}^{*}$ and $\Killing(E_{\alpha},E_{\alpha}^{*})=-\Killing(E_{\alpha},E_{-\alpha})=1$
for all $\alpha\in\Roots$ (\term{Weyl's canonical basis}\index{zzz@Weyl's canonical basis}\hypertarget{Weyl_27s+canonical+basis}{}). %
Then $\{E_{\alpha},T_{i},E_{\alpha}^{*}|\ \alpha\in\Roots^{+}\}$
is a basis of $\fg_{\C}$ with dual basis $\{E_{\alpha}^{*},T_{i},E_{\alpha}|\ \alpha\in\Roots^{+}\}$
w.r.t. $\Killing$. 

The left and right regular representation $\symb{\Ltrans}\index{${TL}$@${\Ltrans}$}\hypertarget{ref.36}{}$ and
$\symb{\Rtrans}\index{${TR}$@${\Rtrans}$}\hypertarget{ref.37}{}$ of $\rG$ on $L^{2}(\rG)$ are defined by
\[
\left(\Ltrans(g)f\right)(x):=f(g^{-1}x),\quad\left(\Rtrans(g)f\right)(x):=f(xg),\qquad g\in\rG,\ f\in L^{2}(\rG)\ 
\]
For $X\in\fg$, let $\symb{X^{\Lft}}\index{${XR}$@${X^{\Lft}}$}\hypertarget{ref.38}{}:=\d\Ltrans(X)$ and $\symb{X^{\Rght}}\index{${XL}$@${X^{\Rght}}$}\hypertarget{ref.39}{}:=\d\Rtrans(X)$.
That is, $X^{\Rght}$ and $X^{\Lft}$ are the differential operators
on $C^{\infty}(\rG)$ defined by

\[
\left(\symb{X^{\Lft}}\index{${XR}$@${X^{\Lft}}$}\hypertarget{ref.40}{}f\right)(g):=\frac{\d}{\d t}f(e^{-tX}g)\big|_{t=0},\qquad\left(\symb{X^{\Rght}}\index{${XR}$@${X^{\Rght}}$}\hypertarget{ref.41}{}f\right)(g):=\frac{\d}{\d t}f(ge^{tX})\big|_{t=0},
\]
for $g\in\rG,\ f\in C^{\infty}(\rG)$. For $Z=X+\im Y\in\fg_{\C}$
with $X,Y\in\fg$, let
\[
\symb{Z^{\Lft}}\index{${ZR}$@${Z^{\Lft}}$}\hypertarget{ref.42}{}:=X^{\Lft}+\im Y^{\Lft},\quad\symb{Z^{\Rght}}\index{${ZR}$@${Z^{\Rght}}$}\hypertarget{ref.43}{}:=X^{\Rght}+\im Y^{\Rght}.
\]
Let $\symb{\scU(\fg_{\C})}\index{${U(g)}$@${\scU(\fg_{\C})}$}\hypertarget{ref.44}{}$ be the universal enveloping algebra
of $\fg_{\C}$. The maps $Z\mapsto Z^{\Lft}$ and $Z\mapsto Z^{\Rght}$
are representations of $\fg_{\C}$ on $C^{\infty}(\rG)$, and hence
the definitions of $Z^{\Lft}$ and $Z^{\Rght}$ are naturally extended
for all $Z\in\scU(\fg_{\C})$.

Define $\phi_{\lambda}:\rG\to\C$ by 
\[
\symb{\phi_{\lambda}}\index{${philam}$@${\phi_{\lambda}}$}\hypertarget{ref.45}{}(g):=\left\langle \hwvec|\pi_{\lambda}(g)\hwvec\right\rangle ,\qquad g\in\rG.
\]
We see $\pi_{\lambda}(e^{X})\hwvec=e^{\lambda(X)}\hwvec$ for $X\in\frakt$,
and hence $\phi_{\lambda}(e^{X})=e^{\lambda(X)}$.

Define the subspace $\symb{\cH_{\lambda}(\rG)}\index{${Hlam(G)}$@${\cH_{\lambda}(\rG)}$}\hypertarget{ref.46}{}\subset C^{\infty}(\rG)$
to be the set of $f\in C^{\infty}(\rG)$ such that
\begin{align}
X^{\Rght}f=0,\ \forall X\in\frakn^{-}\quad\text{and}\quad f(gt)=\phi_{\lambda}(t)^{-1}f(g),\ \forall t\in\rT,\forall g\in\rG.\ \label{eq:def:Hlamb(G)}\displabel{eq:def:Hlamb(G)}
\end{align}
Note that $X^{\Rght}f=0$ for all $X\in\frakn^{-}$ if and only if
$\left(E_{\alpha}^{*}\right)^{\Rght}f=0$ for all $\alpha\in\Roots^{+}.$

The \term{Borel--Weil theorem}\index{zzz@Borel--Weil theorem}\hypertarget{Borel--Weil+theorem}{}, which is a complex-geometric representation
of the irreducible unitary representations of $\rG$, is proven in
two ways: analytically or algebraically. (For a concise exposition
of the Borel--Weil theorem, see e.g. \cite{AlF10}.) The analytic
proof begins with the Cartan--Weyl highest weight theory and the Peter--Weyl
theorem, and it is completed via the following ``\term{pre-Borel--Weil theorem}\index{zzz@pre-Borel--Weil theorem}\hypertarget{pre-Borel--Weil+theorem}{}'':

\begin{shaded}%
\begin{thm}
\label{thm:pre-Borel-Weil}\displabel{thm:pre{-}Borel{-}Weil} (i) $\cH_{\lambda}(\rG)$ is invariant
under $\Ltrans(\rG)$;

(ii) $g\mapsto\Ltrans(g)|_{\cH_{\lambda}(\rG)}$ is an irreducible
unitary representation of $\rG$ on $\cH_{\lambda}(\rG)\subset L^{2}(\rG)$
with highest weight $\lambda$. \end{thm}
\end{shaded}

In this paper, we do not use any complex-geometric method including
the Borel--Weil theorem, but use the above seemingly non-geometric
statement of the pre-Borel--Weil theorem.

\section{Casimir and Laplacian}

\label{sec:Casimir}\displabel{sec:Casimir}

Define $\Casi_{\pm},\Casi_{0}\in\scU(\fg_{\C})$ by

\[
\symb{\Casi_{-}}\index{${Casi0}$@${\Casi_{-}}$}\hypertarget{ref.47}{}:=\sum_{\alpha\in\Roots^{+}}E_{\alpha}^{*}E_{\alpha},\quad\symb{\Casi_{+}}\index{${Casi0}$@${\Casi_{+}}$}\hypertarget{ref.48}{}:=\sum_{\alpha\in\Roots^{+}}E_{\alpha}E_{\alpha}^{*},\quad\symb{\Casi_{0}}\index{${Casi0}$@${\Casi_{0}}$}\hypertarget{ref.49}{}:=\sum_{i=1}^{\ell}T_{i}^{2}.
\]
Recall $\rho$ is the half sum of positive roots. Then we see
\[
\Casi_{+}-\Casi_{-}=2\nu^{-1}(\rho),\ 
\]
The \term{Casimir element}\index{zzz@Casimir element}\hypertarget{Casimir+element}{} $\Casi\in\scU(\fg_{\C})$ is defined
by
\begin{align}
\symb{\Casi}\index{${Casi}$@${\Casi}$}\hypertarget{ref.50}{} & :=\Casi_{0}+\Casi_{+}+\Casi_{-}=\Casi_{0}+2\nu^{-1}(\rho)+2\Casi_{-}\nonumber \\
 & =\Casi_{0}+2\nu^{-1}(\rho)+2\left(\Casi_{+}-2\nu^{-1}(\rho)\right)=\Casi_{0}-2\nu^{-1}(\rho)+2\Casi_{+}.\label{eq:def:Casimir}\displabel{eq:def:Casimir}
\end{align}
Define the \term{Laplacian}\index{zzz@Laplacian}\hypertarget{Laplacian}{} $\Delta$ on $\rG$ by
\[
\symb{\Delta}\index{${Delta}$@${\Delta}$}\hypertarget{ref.51}{}:=\Casi^{\Rght}=\Casi^{\Lft}.
\]
Let $\{\symb{X_{k}}\index{${Xk}$@${X_{k}}$}\hypertarget{ref.52}{}\}$ be an orthonormal basis of $\fg$, i.e.
$\left\langle X_{k}|X_{l}\right\rangle _{\fg}:=-\Killing(X_{k},X_{l})=\delta_{ij}$.
Then by the basic properties of the Casimir elements, we have 
\[
\Delta=-\sum_{k}\left(X_{k}^{\Rght}\right)^{2}.
\]
By the pre-Borel--Weil theorem \ref{thm:pre-Borel-Weil} , we also
have 
\[
\Delta|_{\cH_{\lambda}(\rG)}=\left(\left(\lambda+\rho,\lambda+\rho\right)-\left(\rho,\rho\right)\right)\Id.
\]
{} %

\section{Magnetic Laplacian}

\label{sec:magnetic}\displabel{sec:magnetic}

Let $M$ be a Riemannian manifold, and $\theta$ be a $\im\R$-valued
1-form on $M$. Define the \term{magnetic exterior differentiation}\index{zzz@magnetic exterior differentiation}\hypertarget{magnetic+exterior+differentiation}{}
$\d^{\theta}:C^{\infty}(M,\C)\to\Lambda^{1}(M,\C)$ by $\symb{\d^{\theta}}\index{${dalph}$@${\d^{\theta}}$}\hypertarget{ref.53}{}:=\d+\theta$,
i.e. $\d^{\theta}f:=\d f+f\theta$ for $f\in C^{\infty}(M,\C)$, and
the \term{magnetic Laplacian}\index{zzz@magnetic Laplacian}\hypertarget{magnetic+Laplacian}{} $\Delta^{\theta}:C^{\infty}(M,\C)\to C^{\infty}(M,\C)$
by
\[
\symb{\Delta^{\theta}}\index{${Deltath}$@${\Delta^{\theta}}$}\hypertarget{ref.54}{}:=(\d^{\theta})^{*}\d^{\theta}.
\]
where $(\d^{\theta})^{*}$ is the formal adjoint of $\d^{\theta}$
with respect to the $L^{2}$-inner product of functions and 1-forms.
Note that $\im\R=\fraku(1)$ (the Lie algebra of ${\rm U}(1)$), and
hence $\d^{\theta}$ can be viewed as a covariant derivative on the
trivial line bundle $M\times\C$, associated with the trivial ${\rm U}(1)$-principal
bundle $M\times{\rm U}(1)$. Thus $\Delta^{\theta}$ is nothing but
the Bochner Laplacian corresponding to this covariant derivative.
For further information on magnetic Laplacians on manifolds, see e.g.
\cite{Shu01,ELMP16}.

The Lie group $\rG$ has a Riemannian metric given by the inner product
on $\fg$: $\left\langle X|Y\right\rangle _{\fg}:=-\Killing(X,Y)$.
For $\alpha\in\im\fg^{*}\subset\fg_{\C}^{*}$, let $\alpha^{\Rght}$
be the $\im\R$-valued 1-form on $\rG$ defined by (\ref{eq:def:alphaR}),
and define $\d^{\alpha}:C^{\infty}(\rG,\C)\to\Lambda^{1}(\rG,\C)$
and $\Delta^{\alpha}:C^{\infty}(\rG,\C)\to C^{\infty}(\rG,\C)$ by

\[
\symb{\d^{\alpha}}\index{${dalp}$@${\d^{\alpha}}$}\hypertarget{ref.55}{}\equiv\symb{\d_{\Rght}^{\alpha}}\index{${dRalp}$@${\d_{\Rght}^{\alpha}}$}\hypertarget{ref.56}{}:=\d^{\alpha^{\Rght}},\qquad\symb{\Delta^{\alpha}}\index{${Deltaalp}$@${\Delta^{\alpha}}$}\hypertarget{ref.57}{}\equiv\symb{\Delta_{\Rght}^{\alpha}}\index{${DeltaRalp}$@${\Delta_{\Rght}^{\alpha}}$}\hypertarget{ref.58}{}:=\left(\d^{\alpha}\right)^{*}\d^{\alpha}=\Delta^{\alpha^{\Rght}},
\]
Let $\{X_{k}\}$ be an orthonormal basis of $\fg$, and %
{} $\symb{\xi_{k}}\index{${xik}$@${\xi_{k}}$}\hypertarget{ref.59}{}:=\nu(X_{k})$. Let $\alpha=\im\sum_{k}a_{k}\xi_{k}\in\im\fg^{*}$
($a_{k}\in\R$). Then we have%

\begin{equation}
\d^{\alpha}f=\sum_{k}\left(X_{k}^{\Rght}f+\im a_{k}f\right)\xi_{k}^{\Rght}\qquad f\in C^{\infty}(\rG)\label{eq:dalphaExplicit}\displabel{eq:dalphaExplicit}
\end{equation}
For a 1-form $A=\sum_{k}A_{k}\xi_{k}^{\Rght}\in\Lambda^{1}(\rG,\R)$
($A_{k}\in C^{\infty}(\rG,\R)$), the adjoint operator $\left(\d^{\alpha}\right)^{*}:\Lambda^{1}(\rG,\R)\to C^{\infty}(\rG)$
is explicitly expressed by

\begin{equation}
\left(\d^{\alpha}\right)^{*}A=-\sum_{k}\left(X_{k}^{\Rght}+\im a_{k}\right)A_{k}.\label{eq:dalpha*Explicit}\displabel{eq:dalpha{*}Explicit}
\end{equation}
and hence $\Delta^{\alpha}$ is written as
\begin{equation}
\Delta^{\alpha}=-\sum_{k}\left(X_{k}^{\Rght}+\im a_{k}\right)^{2}=-\sum_{k}\left(X_{k}^{\Rght}\right)^{2}-2\im\sum_{k}a_{k}X_{k}^{\Rght}+\sum_{k}a_{k}^{2}.\label{eq:DeltaalphExplicit}\displabel{eq:DeltaalphExplicit}
\end{equation}
The inner product $\left\langle \bullet|\bullet\right\rangle _{\fg}$
on $\fg$ is naturally extended to the Hermitian inner product $\left\langle \bullet|\bullet\right\rangle _{\fg_{\C}}$
on $\fg_{\C}$. This induces the natural Hermitian inner product $\left\langle \bullet|\bullet\right\rangle =\left\langle \bullet|\bullet\right\rangle _{\fg_{\C}^{*}}$
on $\fg_{\C}^{*}$. (Although $\left\langle \alpha|\beta\right\rangle =(\alpha,\beta)$
holds for $\alpha,\beta\in\im\fg^{*}$, we prefer the notation $\left\langle \bullet|\bullet\right\rangle $
to $(\bullet,\bullet)$ so as to be more consistent with the Hilbert
space structure of $L^{2}(\rG)$.) Then (\ref{eq:DeltaalphExplicit})
is written as the following coordinate-free form:

\begin{shaded}%
\begin{lem}
Let $\alpha\in\im\fg^{*}$. Then
\begin{equation}
\Delta^{\alpha}=\Delta-2\nu^{-1}(\alpha)^{\Rght}+\left\langle \alpha|\alpha\right\rangle _{\fg_{\C}^{*}}\ \label{eq:Deltaalpha-g}\displabel{eq:Deltaalpha{-}g}
\end{equation}
\end{lem}
\end{shaded}

For $\alpha=\im\sum_{i=1}^{\ell}a_{i}\nu(\im T_{i})\in\im\frakt^{*}$
($a_{i}\in\R$), the \term{$\frakt$-partial magnetic Laplacian}\index{zzz@$\frakt$-partial magnetic Laplacian}\hypertarget{_24_5Cfrakt_24-partial+magnetic+Laplacian}{}
$\symb{\Delta_{\frakt}^{\alpha}}\index{${DeltaTlam}$@${\Delta_{\frakt}^{\alpha}}$}\hypertarget{ref.60}{}$ is defined by restricting
Eq. (\ref{eq:DeltaalphExplicit}) to $\frakt$, i.e.%
\begin{align*}
\symb{\Delta_{\frakt}^{\alpha}}\index{${DeltaTlam}$@${\Delta_{\frakt}^{\alpha}}$}\hypertarget{ref.61}{}: & =-\sum_{i=1}^{\ell}\left(\left(\im T_{i}\right)^{\Rght}+\im a_{i}\right)^{2}=-\sum_{i=1}^{\ell}\left(\left(\im T_{i}\right)^{\Rght}-\alpha(\im T_{i})\right)^{2}\ \\
 & =\sum_{i=1}^{\ell}\left(T_{i}^{\Rght}+a_{i}\right)^{2}=\sum_{i=1}^{\ell}\left(\left(T_{i}\right)^{\Rght}-\alpha(T_{i})\right)^{2}
\end{align*}
Then we have an analogue of (\ref{eq:Deltaalpha-g}):

\begin{equation}
\Delta_{\frakt}^{\alpha}=\Delta_{\frakt}-2\nu^{-1}(\alpha)^{\Rght}+\left\langle \alpha|\alpha\right\rangle _{\fg_{\C}^{*}},\qquad\symb{\Delta_{\frakt}}\index{${Deltat}$@${\Delta_{\frakt}}$}\hypertarget{ref.62}{}:=\Casi_{0}^{\Rght}.\label{eq:Deltatalph-nocoord}\displabel{eq:Deltatalph{-}nocoord}
\end{equation}

\begin{shaded}%
\begin{lem}
For $\lambda\in\im\frakt^{*}$ and the half sum of positive roots
$\rho$, we have
\begin{equation}
\Delta^{-(\lambda+\rho)}=\Delta_{\frakt}^{-\lambda}+2\Casi_{+}^{\Rght}+\left\langle 2\lambda+\rho|\rho\right\rangle _{\fg_{\C}^{*}}.\ \label{eq:Delta-(la+rho)=00003D}\displabel{eq:Delta{-}(la+rho)=00003D}
\end{equation}
\end{lem}
\end{shaded}
\begin{proof}
We have
\begin{align*}
\Delta^{-(\lambda+\rho)} & =_{{\scriptstyle (\ref{eq:Deltaalpha-g})}}\Delta-2\nu^{-1}(-\lambda-\rho)^{\Rght}+\left\langle -\lambda-\rho|-\lambda-\rho\right\rangle _{\fg_{\C}^{*}}\ \\
 & =\Casi^{\Rght}+2\left[\nu^{-1}(\lambda)^{\Rght}+\nu^{-1}(\rho)^{\Rght}\right]+\left\langle \lambda+\rho|\lambda+\rho\right\rangle _{\fg_{\C}^{*}}\\
 & =_{{\scriptstyle (\ref{eq:def:Casimir})}}\left(\Casi_{0}-2\nu^{-1}(\rho)+2\Casi_{+}\right)^{\Rght}+2\nu^{-1}(\lambda)^{\Rght}+2\nu^{-1}(\rho)^{\Rght}+\left\langle \lambda+\rho|\lambda+\rho\right\rangle _{\fg_{\C}^{*}}\\
 & =\Casi_{0}^{\Rght}+2\nu^{-1}\lambda^{\Rght}+\left\langle \lambda|\lambda\right\rangle _{\fg_{\C}^{*}}+2\Casi_{+}^{\Rght}+2\left\langle \lambda|\rho\right\rangle _{\fg_{\C}^{*}}+\left\langle \rho|\rho\right\rangle _{\fg_{\C}^{*}}\ \\
 & =_{{\scriptstyle (\ref{eq:Deltatalph-nocoord})}}\Delta_{\frakt}^{-\lambda}+2\Casi_{+}^{\Rght}+2\left\langle \lambda|\rho\right\rangle _{\fg_{\C}^{*}}+\left\langle \rho|\rho\right\rangle _{\fg_{\C}^{*}}\\
 & =\Delta_{\frakt}^{-\lambda}+2\Casi_{+}^{\Rght}+\left\langle 2\lambda+\rho|\rho\right\rangle _{\fg_{\C}^{*}}.\ \ 
\end{align*}

\end{proof}

\begin{shaded}%
\begin{lem}
Let $f\in C^{\infty}(\rG)$. Then %
\begin{equation}
f(gt)=\phi_{\lambda}(t)^{-1}f(g)\text{ for all }t\in\rT,\ g\in\rG\quad\text{if and only if }\quad\Delta_{\frakt}^{-\lambda}f=0.\label{eq:Deltatlamf=00003D0}\displabel{eq:Deltatlamf=00003D0}
\end{equation}
\end{lem}
\end{shaded}
\begin{proof}
We see that $f(gt)=\phi_{\lambda}(t)^{-1}f(g),\ \forall t\in\rT,\forall g\in\rG$
if and only if 
\begin{align*}
\forall X & \in\frakt,\forall g\in\rG,\quad\frac{\d}{\d\epsilon}f(ge^{\epsilon X})\Big|_{\epsilon=0}=\frac{\d}{\d\epsilon}\phi_{\lambda}(e^{\epsilon X})^{-1}f(g)\Big|_{\epsilon=0},\\
\text{iff } & \forall X\in\frakt,\ \left(X^{\Rght}+\lambda(X)\right)f=0,\\
\text{iff } & \forall i,\ \left(\left(\im T_{i}\right)^{\Rght}+\lambda(\im T_{i})\right)f=0,\\
\text{iff } & -\sum_{i}\left(\left(\im T_{i}\right)^{\Rght}+\lambda(\im T_{i})\right)^{2}f\equiv\Delta_{\frakt}^{-\lambda}f=0.
\end{align*}

\end{proof}
For any linear operator $A$, let $\symb{\spec}\index{${spec}$@${\spec}$}\hypertarget{ref.63}{}A$ denote the
spectrum of $A$. 

\begin{shaded}%
\begin{thm}
(1) Let $c_{\lambda}:=\inf{\rm spec}\Delta^{-(\lambda+\rho)}$. Then
\[
c_{\lambda}=\left\langle 2\lambda+\rho|\rho\right\rangle _{\fg_{\C}^{*}}.
\]
 (2) $\cH_{\lambda}(\rG)$ is the ``ground eigenspace'' of $\Delta^{-(\lambda+\rho)}$,
i.e. 
\[
\cH_{\lambda}(\rG)=\ker\left[\Delta^{-(\lambda+\rho)}-c_{\lambda}\right]
\]

\end{thm}
\end{shaded}
\begin{proof}
Since $\Delta_{\frakt}^{-\lambda}$ and $\Casi_{+}^{\Rght}$ are positive
semidefinite operators, we find by (\ref{eq:Delta-(la+rho)=00003D}),
\[
\Delta^{-(\lambda+\rho)}-\left\langle 2\lambda+\rho|\rho\right\rangle _{\fg_{\C}^{*}}=\Delta_{\frakt}^{-\lambda}+2\Casi_{+}^{\Rght}\ge O.\ 
\]
By (\ref{eq:Deltatlamf=00003D0}) and
\[
\forall X\in\frakn^{-},\ X^{\Rght}f=0\iff\forall\alpha\in\Roots^{+},\ \left(E_{\alpha}^{*}\right)^{\Rght}f=0\iff2\Casi_{+}^{\Rght}f=0\ 
\]
we have
\begin{align*}
\cH_{\lambda}(\rG) & =\left\{ f\in C^{\infty}(G):\ 2\Casi_{+}^{\Rght}f=0\ \&\ \Delta_{\frakt}^{-\lambda}f=0\right\} \\
 & =\left\{ f\in C^{\infty}(G):\ \left[2\Casi_{+}^{\Rght}+\Delta_{\frakt}^{-\lambda}\right]f=0\right\} \\
 & =\left\{ f\in C^{\infty}(G):\ \left(\Delta^{-(\lambda+\rho)}-\left\langle 2\lambda+\rho|\rho\right\rangle _{\fg_{\C}^{*}}\right)f=0\right\} \\
 & =\ker\left[\Delta^{-(\lambda+\rho)}-\left\langle 2\lambda+\rho|\rho\right\rangle _{\fg_{\C}^{*}}\right].
\end{align*}

\end{proof}

\section{GS quantization on $\protect\cH_{\lambda}(\protect\rG)$}

\label{sec:GSquantOnH}\displabel{sec:GSquantOnH}

In Sec. \ref{sec:mainTheorem}, we defined the GS quantization $\cQ$
for an irreducible unitary representation $(\pi_{\lambda},V_{\lambda})$
with highest weight $\lambda\in\cX_{+}(\rT)\subset\im\frakt$. In
the following, we set $(\pi_{\lambda},V_{\lambda})=(\Ltrans,\cH_{\lambda}(\rG))$,
and examine the GS quantization there.

Let $\hwvec\in\cH_{\lambda}(\rG)\subset L^{2}(\rG)$ ($\left\Vert \hwvec\right\Vert =1$)
be the highest weight vector, i.e. $X^{\Lft}\hwvec=\lambda(X)\hwvec,\ \forall X\in\frakt$,
such that $\hwvec(1_{\rG})>0$. For $u,v\in L^{2}(\rG)$, define $\Lft_{u,v}\in C(\rG)$
by

\[
\symb{\Lft_{u,v}(g)}\index{${Luv}$@${\Lft_{u,v}(g)}$}\hypertarget{ref.64}{}:=\left\langle u|\Ltrans(g)v\right\rangle ,\qquad\symb{\Rght_{u,v}(g)}\index{${Ruv}$@${\Rght_{u,v}(g)}$}\hypertarget{ref.65}{}:=\left\langle u|\Rtrans(g)v\right\rangle .
\]
Recall $\symb{\phi_{\lambda}}\index{${philamb}$@${\phi_{\lambda}}$}\hypertarget{ref.66}{}(g):=\left\langle \hwvec|\Ltrans(g)\hwvec\right\rangle =\Lft_{\hwvec,\hwvec}(g)$.

\begin{shaded}%
\begin{lem}
For any $X\in\fg_{\C}$ and $u,v\in L^{2}(\rG)$,
\begin{equation}
X^{\Rght}\Lft_{u,v}=\Lft_{u,X^{\Lft}v},\qquad X^{\Lft}\Lft_{u,v}=\Lft_{-\left(X^{*}\right)^{\Lft}u,\, v},\label{eq:XRL,XLL}\displabel{eq:XRL,XLL}
\end{equation}
\begin{equation}
X^{\Rght}\ol{\Lft_{u,v}}=\ol{\Lft_{u,\,-\left(X^{*}\right)^{\Lft}v}},\qquad X^{\Lft}\ol{\Lft_{u,v}}=\ol{\Lft_{X^{\Lft}u,v}}.\label{eq:XRLbar,XLLbar}\displabel{eq:XRLbar,XLLbar}
\end{equation}
\end{lem}
\end{shaded}
\begin{proof}
For $X\in\fg$, we have
\begin{align*}
\left(X^{\Rght}\Lft_{u,v}\right)(g) & =\frac{\d}{\d t}\Lft_{u,v}(ge^{tX})|_{t=0}=\frac{\d}{\d t}\left\langle u|\Ltrans(ge^{tX})v\right\rangle |_{t=0}\ \\
 & =\frac{\d}{\d t}\left\langle u|\Ltrans(g)\Ltrans(e^{tX})v\right\rangle |_{t=0}=\left\langle u|\Ltrans(g)X^{\Lft}v\right\rangle =\Lft_{u,X^{\Lft}v}(g).\ 
\end{align*}
Hence, for $Z=X+\im Y\in\fg_{\C}$ with $X,Y\in\fg$, we have
\[
Z^{\Rght}\Lft_{u,v}=X^{\Rght}\Lft_{u,v}+\im Y^{\Rght}\Lft_{u,v}=\Lft_{u,X^{\Lft}v}+\im\Lft_{u,Y^{\Lft}v}=\Lft_{u,X^{\Lft}v+\im Y^{\Lft}v}=\Lft_{u,Z^{\Lft}v}.\ 
\]
Other relations are shown similarly.
\end{proof}
\begin{shaded}%
\begin{lem}
\label{thm:(TR,TRphi),(TL,TLphibar)}\displabel{thm:(TR,TRphi),(TL,TLphibar)} (1) $\left(\Rtrans,{\rm span}\left\{ \Rtrans(\rG)\phi_{\lambda}\right\} \right)$
and $\left(\Ltrans,{\rm span}\left\{ \Ltrans(\rG)\ol{\phi_{\lambda}}\right\} \right)$
are irreducible unitary representations of $\rG$ with the highest
weight $\lambda$, where $\phi_{\lambda}$ and $\ol{\phi_{\lambda}}$
are highest weight vectors, respectively.

(2) $\left(\Ltrans,{\rm span}\left\{ \Ltrans(\rG)\phi_{\lambda}\right\} \right)$
and $\left(\Rtrans,{\rm span}\left\{ \Rtrans(\rG)\ol{\phi_{\lambda}}\right\} \right)$
are irreducible unitary representations of $\rG$ with the lowest
weight $-\lambda$, where $\phi_{\lambda}$ and $\ol{\phi_{\lambda}}$
are lowest weight vectors, respectively.\end{lem}
\end{shaded}
\begin{proof}
Since $\hwvec$ is the highest weight vector of $\left(\Ltrans,\cH_{\lambda}(\rG)\right)$,
we have 
\[
\forall\alpha\in\Roots^{+},\quad E_{\alpha}^{\Lft}\hwvec=0
\]
By (\ref{eq:XRL,XLL}), we have
\[
E_{\alpha}^{\Rght}\phi_{\lambda}=E_{\alpha}^{\Rght}\Lft_{\hwvec,\hwvec}=\Lft_{\hwvec,E_{\alpha}^{\Lft}\hwvec}=\Lft_{\hwvec,0}=0.
\]
Hence $\phi_{\lambda}$ is the highest weight vector of $\left(\Rtrans,{\rm span}\left\{ \Rtrans(\rG)\phi_{\lambda}\right\} \right)$.
By (\ref{eq:XRLbar,XLLbar}),
\[
E_{\alpha}^{\Lft}\ol{\phi_{\lambda}}=\ol{E_{\alpha}^{\Lft}\phi_{\lambda}}=\ol{E_{\alpha}^{\Lft}\Lft_{\hwvec,\hwvec}}=\ol{\Lft_{E_{\alpha}^{\Lft}\hwvec,\hwvec}}=\ol{\Lft_{0,\hwvec}}=0.\ 
\]
Hence $\ol{\phi_{\lambda}}$ is the highest weight of $\left(\Ltrans,{\rm span}\left\{ \Ltrans(\rG)\ol{\phi_{\lambda}}\right\} \right)$.
The proof of (2) is similar.
\end{proof}
\begin{shaded}%
\begin{lem}
\label{thm:vlam{eq}dphibar}\displabel{thm:vlam\{eq\}dphibar}We have $\hwvec=d_{\lambda}^{1/2}\ol{\phi_{\lambda}}$.%
\end{lem}
\end{shaded}
\begin{proof}
By Lemma \ref{thm:(TR,TRphi),(TL,TLphibar)}, $u:=\ol{\phi_{\lambda}}$
is the highest weight vector of $\left(\Ltrans,{\rm span}\left\{ \Ltrans(\rG)u\right\} \right)$,
and also the lowest weight vector of $\left(\Rtrans,{\rm span}\left\{ \Rtrans(\rG)u\right\} \right)$.
Such $u\in L^{2}(\rG)$ is unique up to scalar multiple by the Peter--Weyl
theorem. If we set $u:=\hwvec$, we have the same statement, since
$\cH_{\lambda}(\rG)={\rm span}\left\{ \Ltrans(\rG)\hwvec\right\} $.
Hence $\hwvec=z\ol{\phi_{\lambda}}$ for some $z\in\C\setminus\{0\}$.
We see $\left\Vert \ol{\phi_{\lambda}}\right\Vert =d_{\lambda}^{-1/2}$
and $\ol{\phi_{\lambda}}(1_{\rG})=1$. Thus, recalling that we are
assuming $\hwvec(1_{\rG})>0$, we have $z=d_{\lambda}^{1/2}$.
\end{proof}
The following easily shown lemma will not used later, but it will
help to understand the relation to the notion of \emph{reproducing
kernel}. (See e.g. \cite{Lan98} for the relations between coherent
states, quantizations and reproducing kernels; see \cite{Nee10} for
unitary representation theory in terms of reproducing kernels.)

\begin{shaded}%
\begin{lem}
[reproducing kernel] For $g\in\rG$, define the $\cH_{\lambda}(\rG)$-\term{delta function}\index{zzz@delta function}\hypertarget{delta+function}{}
$\delta_{\lambda,g}\in\cH_{\lambda}(\rG)$ by 
\[
\symb{\delta_{\lambda,g}}\index{${deltalamg}$@${\delta_{\lambda,g}}$}\hypertarget{ref.67}{}:=d_{\lambda}\Ltrans(g)\ol{\phi_{\lambda}}=d_{\lambda}^{1/2}\Ltrans(g)\hwvec=d_{\lambda}\ol{\Lft_{\Ltrans(g)\hwvec,\hwvec}}.
\]
Then
\[
v(g)=\left\langle \delta_{\lambda,g}|v\right\rangle ,\qquad\forall v\in\cH_{\lambda}(\rG),\ \forall g\in\rG.
\]
$K(g,h):=\delta_{\lambda,1_{\rG}}(h^{-1}g)$ is called the $\cH_{\lambda}(\rG)$-\term{reproducing kernel}\index{zzz@reproducing kernel}\hypertarget{reproducing+kernel}{}.\end{lem}
\end{shaded}

Let $\Hwvproj:=\hwvec\hwvec^{*}$. We view $\Hwvproj$ \emph{not}
as a projection from $V_{\lambda}=\cH_{\lambda}(\rG)$ onto $\C\hwvec$,
but as a projection from $L^{2}(\rG)$ onto $\C\hwvec$. (Both views
are consistent by the orthogonality relations in the Peter--Weyl theory.)
Let

\[
\symb{g\cdot\Hwvproj}\index{${g.E}$@${g\cdot\Hwvproj}$}\hypertarget{ref.68}{}\equiv\symb{\Hwvproj(g)}\index{${Elamb(g)}$@${\Hwvproj(g)}$}\hypertarget{ref.69}{}:=\Ltrans(g)\Hwvproj\Ltrans(g^{-1}),\qquad g\in\rG.
\]
For $f\in C^{\infty}(\rG,\C)$, define $f(\Hwvproj)\equiv\Hwvproj(f)\in\End(\cH_{\lambda}(\rG))$
by
\[
\symb{f(\Hwvproj)}\index{${f(\Hwvproj)}$@${f(\Hwvproj)}$}\hypertarget{ref.70}{}\equiv\symb{\Hwvproj(f)}\index{${Elamb(f)}$@${\Hwvproj(f)}$}\hypertarget{ref.71}{}:=\int_{\rG}f(g)\Hwvproj(g)\d g.
\]
Precisely, let

\[
\dom\left(\Hwvproj(f)\right):=\left\{ v\in L^{2}(\rG):\ \int_{\rG}\left\Vert f(g)\Hwvproj(g)v\right\Vert \d g<\infty\right\} 
\]
and for each $v\in\dom\left(\Hwvproj(f)\right)$, let%

\[
\Hwvproj(f)v:=\int_{\rG}f(g)\Hwvproj(g)v\d g\qquad\text{(Bochner integral)},
\]
Since we are assuming that $\rG$ is compact and $f\in C^{\infty}(\rG,\C)$
here, we see $\dom\left(\Hwvproj(f)\right)=L^{2}(\rG)$ and that $\Hwvproj(f)$
is a bounded operator.

Note that the definition of $\Hwvproj(f)$ is naturally extended for
any $f\in C^{-\infty}(\rG,\C)$, the space of Schwartz distributions,
since $\cH_{\lambda}(\rG)\subset C^{\infty}(\rG)$ and $\dim\cH_{\lambda}(\rG)<\infty$.

\begin{shaded}%
\begin{lem}
Let $f\in C(\rG,\R)$. Then for any $v_{1},v_{2}\in\cH_{\lambda}(\rG),$
\[
\left\langle v_{1}|\Hwvproj(f)v_{2}\right\rangle =d_{\lambda}^{-1}\left\langle v_{2}|fv_{1}\right\rangle 
\]
where $f$ is regarded as a multiplication operator on $L^{2}(\rG)$
in the rhs.\end{lem}
\end{shaded}
\begin{proof}
Without loss of generality, we can assume that for some $g_{k}\in\rG$,
$k=1,2$, 
\[
v_{k}=\Ltrans(g_{k})\hwvec.
\]
Then we have
\begin{align*}
\left\langle v_{1}|\Hwvproj(f)v_{2}\right\rangle  & =\left\langle \Ltrans(g_{1})\hwvec|\int_{G}f(g)\Hwvproj(g)\d g\Ltrans(g_{2})\hwvec\right\rangle \\
 & =\int_{G}f(g)\left\langle \Ltrans(g_{1})\hwvec|\Ltrans(g)\hwvec\hwvec^{*}\Ltrans(g^{-1})\Ltrans(g_{2})\hwvec\right\rangle \d g\\
 & =\int_{G}f(g)\left\langle \hwvec|\Ltrans(g_{1}^{-1}g)\hwvec\right\rangle \ol{\left\langle \hwvec|\Ltrans(g_{2}^{-1}g)\hwvec\right\rangle }\d g\\
 & =\int_{G}f(g)\left(\Ltrans(g_{1})\phi_{\lambda}\right)(g)\ol{\left(\Ltrans(g_{2})\phi_{\lambda}\right)(g)}\d g\\
 & =\left\langle \Ltrans(g_{1})\phi_{\lambda}|f\Ltrans(g_{2})\phi_{\lambda}\right\rangle =\left\langle \Ltrans(g_{2})\ol{\phi_{\lambda}}|f\Ltrans(g_{1})\ol{\phi_{\lambda}}\right\rangle \\
 & =_{{\scriptstyle \text{Lemma }\ref{thm:vlam{eq}dphibar}}}\left\langle \Ltrans(g_{2})d_{\lambda}^{-1/2}\hwvec|f\Ltrans(g_{1})d_{\lambda}^{-1/2}\hwvec\right\rangle \\
 & =d_{\lambda}^{-1}\left\langle \Ltrans(g_{2})\hwvec|f\Ltrans(g_{1})\hwvec\right\rangle =d_{\lambda}^{-1}\left\langle v_{2}|fv_{1}\right\rangle .\ \ 
\end{align*}

\end{proof}
For $h\in C^{\infty}(\rG\cdot\Hwvproj,\R)$, define $\fextend h\in C^{\infty}(\rG,\R)$
by $\symb{\fextend h}\index{${hhat}$@${\fextend h}$}\hypertarget{ref.72}{}(g):=h(g\cdot\Hwvproj)$, and define the
operator $\cQ(h)$ on $L^{2}(\rG)$, called the \term{Glauber--Sudarshan-type quantization}\index{zzz@Glauber--Sudarshan-type quantization}\hypertarget{Glauber--Sudarshan-type+quantization.1}{}
(or simply, the \term{GS quantization}\index{zzz@GS quantization}\hypertarget{GS+quantization.1}{}), by 

\[
\symb{\cQ(h)}\index{${Q(f)}$@${\cQ(h)}$}\hypertarget{ref.73}{}:=d_{\lambda}\Hwvproj(\fextend h)=d_{\lambda}\int_{G}h(g\cdot\Hwvproj)\Hwvproj(g)\d g,
\]
The following theorem directly follows from the above lemma.

\begin{shaded}%
\begin{thm}
[GS quantization as projection]\label{thm:GSProjection-a}\displabel{thm:GSProjection{-}a} Let $f\in C^{\infty}(\rG,\R)$.
Let $\symb{P_{\lambda}}\index{${Plam}$@${P_{\lambda}}$}\hypertarget{ref.74}{}$ be the orthogonal projection from $L^{2}(\rG)$
onto $\cH_{\lambda}(\rG)$. Then
\[
\Hwvproj(f)=d_{\lambda}^{-1}P_{\lambda}fP_{\lambda}.
\]
where $f$ is regarded as a multiplication operator on $L^{2}(\rG)$
in the rhs. For an orbit function $h\in C^{\infty}(\rG\cdot\Hwvproj,\R)$,
we have $\cQ(h)=P_{\lambda}\fextend hP_{\lambda}$, i.e. 
\[
\cQ(h)v=P_{\lambda}\fextend hv\quad\text{for }v\in\cH_{\lambda}(\rG),\quad\text{and}\quad\cQ(h)v=0\quad\text{for }v\in\cH_{\lambda}(\rG)^{\perp}.
\]
\end{thm}
\end{shaded}

\section{Asymptotic representation}

\label{sec:asymp}\displabel{sec:asymp}

Let
\[
\symb{\ul{\Delta}^{-(\lambda+\rho)}}\index{${Deltaub}$@${\ul{\Delta}^{-(\lambda+\rho)}}$}\hypertarget{ref.75}{}:=\Delta^{-(\lambda+\rho)}-\inf{\rm spec}\Delta^{-(\lambda+\rho)}=\Delta^{-(\lambda+\rho)}-\left\langle 2\lambda+\rho|\rho\right\rangle _{\fg_{\C}^{*}}.
\]
Let $V\in C^{\infty}(\rG,\R)$. For $\ss>0$, define the operator
$T_{\ss}$ by
\[
\symb{T_{\ss}}\index{${Tnun}$@${T_{\ss}}$}\hypertarget{ref.76}{}:=\ss\ul{\Delta}^{-(\lambda+\rho)}+\im V
\]
Then $T_{\ss}$ is a closed operator satisfying
\[
\Re\langle v|T_{\ss}v\rangle\ge0
\]
for all $v\in\dom(T_{\ss})=\dom(\ul{\Delta}^{-(\lambda+\rho)})$.
Hence $T_{\ss}$ generates the strongly continuous contraction semigroup
$\{e^{-tT_{\ss}}|t\ge0\}$ by the Hille--Yosida Theorem \cite{RS75}. 

Note that $\ul{\Delta}^{-(\lambda+\rho)}$ is a compact operator on
$L^{2}(\rG)$. %
{} Hence we have the spectrum decomposition

\[
\ul{\Delta}^{-(\lambda+\rho)}=\sum_{k=0}^{\infty}\eigenval_{k}E_{k},\quad0=\eigenval_{0}<\eigenval_{1}<\cdots
\]
where each $E_{k}$ is an orthogonal projection, and $\sum_{k}E_{k}=I$.

If $V\in C^{\infty}(\rG,\R)$ and $f\in C^{\infty}(\rG,\C)$, we see
{} $\symb{f_{\ss,t}}\index{${fs,t}$@${f_{\ss,t}}$}\hypertarget{ref.77}{}:=e^{-tT_{\ss}}f\in C^{\infty}(\rG,\C)$ and
$\symb{g_{\ss,t}}\index{${gsig,t}$@${g_{\ss,t}}$}\hypertarget{ref.78}{}:=P_{\lambda}e^{-tT_{\ss}}f=P_{\lambda}f_{\ss,t}$
for all $t\ge0$.

\begin{shaded}%
\begin{lem}
\label{thm17.2}\displabel{thm17.2}Let $f\in C^{\infty}(\rG)$, $f_{\ss,t}:=e^{-tT_{\ss}}f$
and %
{} $\symb{\eta(t)}\index{${eta(t)}$@${\eta(t)}$}\hypertarget{ref.79}{}:=\left\Vert (I-P_{\lambda})f_{\ss,t}\right\Vert $
then
\begin{equation}
\frac{\d}{\d t}\eta(t)^{2}\le-2\eigenval_{1}\ss\eta(t)^{2}-2\left\Vert Vf_{\ss,t}\right\Vert \eta(t)\label{eq:0518.1}\displabel{eq:0518.1}
\end{equation}

\end{lem}
\end{shaded}
\begin{proof}
Let $\symb{A}\index{${A}$@${A}$}\hypertarget{ref.80}{}:=\ul{\Delta}^{-(\lambda+\rho)}$. Then we easily find
\[
\frac{\d}{\d t}\eta(t)^{2}=\frac{\d}{\d t}\left\Vert \left(1-P_{\lambda}\right)e^{-tT_{\ss}}f\right\Vert ^{2}=-2\ss\left\langle f_{\ss,t}|Af_{\ss,t}\right\rangle -2\Im\left\langle Vf_{\ss,t}|\left(1-P_{\lambda}\right)f_{\ss,t}\right\rangle .
\]
Since $0\le\eigenval_{1}(I-P_{\lambda})\le A$, we have%
\begin{align*}
\frac{\d}{\d t}\eta(t)^{2} & =\frac{\d}{\d t}\left\Vert \left(1-P_{\lambda}\right)e^{-tT_{\ss}}f\right\Vert ^{2}\\
 & \le-2\ss\left\langle f_{\ss,t}|\eigenval_{1}(I-P_{\lambda})f_{\ss,t}\right\rangle -2\Im\left\langle Vf_{\ss,t}|\left(1-P_{\lambda}\right)f_{\ss,t}\right\rangle \\
 & =-2\eigenval_{1}\ss\left\Vert (I-P_{\lambda})f_{\ss,t}\right\Vert ^{2}-2\Im\left\langle Vf_{\ss,t}|\left(1-P_{\lambda}\right)f_{\ss,t}\right\rangle \\
 & \le-2\eigenval_{1}\ss\left\Vert (I-P_{\lambda})f_{\ss,t}\right\Vert ^{2}+2\left\Vert Vf_{\ss,t}\right\Vert \left\Vert \left(1-P_{\lambda}\right)f_{\ss,t}\right\Vert \\
 & =-2\eigenval_{1}\ss\eta(t)^{2}-2\left\Vert Vf_{\ss,t}\right\Vert \eta(t).
\end{align*}

\end{proof}
\begin{shaded}%
\begin{lem}
Suppose $f\in\ker\ul{\Delta}^{-(\lambda+\rho)}\ (=\cH_{\lambda}(\rG))$.
{} Then
\begin{equation}
\forall t>0,\quad\left\Vert f_{\ss,t}-g_{\ss,t}\right\Vert \le\frac{\left\Vert V\right\Vert _{\infty}\left\Vert f\right\Vert }{\eigenval_{1}\ss}\label{eq:|(1-P)e-tTf|<}\displabel{eq:|(1{-}P)e{-}tTf|<}
\end{equation}
\end{lem}
\end{shaded}
\begin{proof}
Recall $\ker\ul{\Delta}^{-(\lambda+\rho)}=\ker(I-P_{\lambda})$ and
$\left\Vert (I-P_{\lambda})e^{-tT_{\ss}}f\right\Vert =\left\Vert (I-P_{\lambda})f_{\ss,t}\right\Vert =\eta(t)$.
Assume $\frac{\d}{\d t}\eta(t)^{2}\ge0$. Then by (\ref{eq:0518.1})%
, we have
\[
\eta(t)^{2}\le\frac{\left\Vert Vf_{\ss,t}\right\Vert \eta(t)}{\eigenval_{1}\ss}.
\]
This implies
\[
\eta(t)\le\frac{\left\Vert Vf_{\ss,t}\right\Vert }{\eigenval_{1}\ss}\le\frac{\left\Vert V\right\Vert \left\Vert f\right\Vert }{\eigenval_{1}\ss}.
\]
Thus we find that
\[
\frac{\d}{\d t}\eta(t)^{2}\ge0\Longrightarrow\eta(t)\le\frac{\left\Vert V\right\Vert \left\Vert f\right\Vert }{\eigenval_{1}\ss},\qquad\forall t>0.
\]
Since $\eta(0)=0$, it follows that%

\[
\left\Vert f_{\ss,t}-g_{\ss,t}\right\Vert =\left\Vert (I-P_{\lambda})e^{-tT_{\ss}}f\right\Vert =\eta(t)\le\frac{\left\Vert V\right\Vert \left\Vert f\right\Vert }{\eigenval_{1}\ss},\qquad\forall t>0.
\]

\end{proof}
\begin{shaded}%
\begin{lem}
\label{thm17.4}\displabel{thm17.4}%
{} %
We have
\begin{equation}
\left\Vert \frac{\d}{\d t}g_{\ss,t}-\im P_{\lambda}VP_{\lambda}g_{\ss,t}\right\Vert \le\left\Vert V\right\Vert _{\infty}\left\Vert \left(1-P_{\lambda}\right)f_{\ss,t}\right\Vert .\label{eq:17.4}\displabel{eq:17.4}
\end{equation}
\end{lem}
\end{shaded}
\begin{proof}
We see
\begin{align*}
 & \frac{\d}{\d t}g_{\ss,t}+\im P_{\lambda}VP_{\lambda}g_{\ss,t}=P_{\lambda}\frac{\d}{\d t}f_{\ss,t}+\im P_{\lambda}VP_{\lambda}P_{\lambda}f_{\ss,t}=P_{\lambda}\left(-T_{\ss}f_{\ss,t}\right)+\im P_{\lambda}VP_{\lambda}f_{\ss,t}\\
 & =-P_{\lambda}\left(\ss\ul{\Delta}^{-(\lambda+\rho)}+\im V\right)f_{\ss,t}+\im P_{\lambda}VP_{\lambda}f_{\ss,t}=-\im P_{\lambda}V\left(1-P_{\lambda}\right)f_{\ss,t}.
\end{align*}
Hence we have
\begin{align*}
 & \left\Vert \frac{\d}{\d t}g_{\ss,t}-\im P_{\lambda}VP_{\lambda}g_{\ss,t}\right\Vert =\left\Vert P_{\lambda}V\left(1-P_{\lambda}\right)f_{\ss,t}\right\Vert \\
 & \qquad\le\left\Vert V\left(1-P_{\lambda}\right)f_{\ss,t}\right\Vert \le\left\Vert V\right\Vert _{\infty}\left\Vert \left(1-P_{\lambda}\right)f_{\ss,t}\right\Vert .
\end{align*}

\end{proof}

\begin{shaded}%
\begin{prop}
[Asymptotic representation]\label{thm:sibori-G}\displabel{thm:sibori{-}G}%
{} Let $f\in\ker\ul{\Delta}^{-(\lambda+\rho)}=\cH_{\lambda}(\rG)$ and
$V\in C^{\infty}(\rG,\R)$. Then for all $t>0$,
\[
\lim_{\ss\to\infty}e^{-tT_{\ss}(V)}f=e^{\im tP_{\lambda}VP_{\lambda}}f,\qquad\symb{T_{\ss}(V)}\index{${Tsig}$@${T_{\ss}(V)}$}\hypertarget{ref.81}{}:=\ss\ul{\Delta}^{-(\lambda+\rho)}+\im V
\]
Especially, for any classical Hamiltonian $h\in C^{\infty}(\rG\cdot\Hwvproj,\R)$
and $t>0$, we have %
\[
e^{\im t\cQ(h)}f=\lim_{\ss\to\infty}e^{-tT_{\ss}(\fextend h)}f,
\]
where $\cQ(h)$ is the GS quantization of $h$.\end{prop}
\end{shaded}
\begin{proof}
By (\ref{eq:17.4}),
\[
\lim_{\ss\to\infty}\left\Vert \frac{\d}{\d t}g_{\ss,t}-\im P_{\lambda}VP_{\lambda}g_{\ss,t}\right\Vert \le\left\Vert V\right\Vert _{\infty}\lim_{\ss\to\infty}\left\Vert \left(1-P_{\lambda}\right)f_{\ss,t}\right\Vert =0
\]
for each $t>0$. This implies
\[
\lim_{\ss\to\infty}g_{\ss,t}=e^{\im tP_{\lambda}VP_{\lambda}}g_{\ss,0}=e^{\im tP_{\lambda}VP_{\lambda}}P_{\lambda}f_{\ss,0}=e^{\im tP_{\lambda}VP_{\lambda}}f,\qquad\forall t>0
\]
Thus by (\ref{eq:|(1-P)e-tTf|<})%
{} we have
\[
\lim_{\ss\to\infty}f_{\ss,t}=\lim_{\ss\to\infty}g_{\ss,t}=e^{\im tP_{\lambda}VP_{\lambda}}f.
\]

\end{proof}

\section{Path integral: Brownian form}

\label{sec:pathInt-Brown}\displabel{sec:pathInt{-}Brown}

{} In this section we give a Brownian path integral representation of
the one-parameter unitary group $\left\{ e^{\im t\cQ(h)}:\ t\in\R\right\} $
where $\cQ(h)$ is the GS quantization of the `classical' Hamiltonian
$h\in C^{\infty}(\rG\cdot\Hwvproj,\R)$. The main tool used here is
the Feynman-Kac-It\^o formula on a vector bundle on a Riemannian
manifold, formulated by G\"uneysu (2010) \cite{Gun10}. The basics
of the theory of Brownian motion on a manifold are summarized in \cite{Gun10}.
The simplest construction of a Brownian motion on a Riemannian manifold
$M$ will be the one which is based on the Nash embedding $M\hookrightarrow\R^{l}$:

\begin{shaded}%
\begin{thm}
\label{thm:Gun2.31}\displabel{thm:Gun2.31}%
{} Let $M\hookrightarrow\R^{l}$ isometrically for some $l\in\N$ and
let the morphism of smooth vector bundles $A:M\times\R^{l}\to TM$
be given as the orthogonal projection $A(x):\R^{l}\to T_{x}M$ for
any $x\in M$. Let $W$ be a Brownian motion in $\R^{l}$. Then the
maximal solution of the stochastic differential equation 
\[
\d X_{t}=\sum_{j=1}^{l}A_{j}(X_{t})\du W_{t}^{j},\qquad X_{0}=x
\]
is a Brownian motion on $M$ with starting point $x$. Here $\du$
denotes the Stratonovich differential.\end{thm}
\end{shaded}

If $M$ is a compact semisimple Lie group $\rG$ embedded in a matrix
Lie group ${\rm GL}(n,\C)$, we have a simpler characterization: Let
$W$ be a Brownian motion on $\fg$. Then the solution of the left
(resp. right) invariant stochastic differential equation $\d X_{t}=X_{t}\du W_{t}$
(resp. $\d X_{t}=\left(\du W_{t}\right)X_{t}$) is a Brownian motion
on $\rG$. However, in this section we does not need a specific definitions
of a Brownian motion on $M$.

Let $M=(M,g)$ be a geodesically and stochastically complete smooth
connected Riemannian manifold. (Any compact Lie group $M$ satisfies
this condition. See \cite{Gun10}.) Let $\alpha$ be a $\im\R$-valued
smooth 1-form on $M$. %
{} Let $V:M\to\R$ be a locally square integrable potential which is
bounded from below, and

\[
\symb{H(\alpha,V)}\index{${H(\alpha,V)}$@${H(\alpha,V)}$}\hypertarget{ref.82}{}:=\frac{1}{2}(\d+\alpha)^{*}(\d+\alpha)+V=\frac{1}{2}\Delta^{\alpha}+V.\ 
\]
The self-adjoint extension of $H(\alpha,V)$ in $L^{2}(M)$ is denoted
again by $H(\alpha,V)$.

\begin{shaded}%
\begin{thm}
{\rm (Feynman--Kac--It\^o formula on a manifold, G\"uneysu \cite{Gun10}) } 
Let $X$ be a Brownian motion in $\rG$. Then%
\[
e^{-tH(\alpha,V)}f(x)=\Ex\left[e^{\cI_{\alpha,V}}f(X_{t})|X_{0}=x\right]\quad{\rm a.e.}\ x\in M,
\]
where
\[
\symb{\cI_{\alpha,V}}\index{${IalphaV}$@${\cI_{\alpha,V}}$}\hypertarget{ref.83}{}:=\int_{0}^{t}\alpha(\du X_{s})-\int_{0}^{t}V(X_{s})\d s.
\]
Here, $\int_{0}^{t}\alpha(\du X_{s})$ stands for the Stratonovich
line integral of $\alpha$ along $B$.\end{thm}
\end{shaded}

This theorem concerns only the cases where $V$ is real-valued. However,
if we confine ourselves to the cases where $\left|V\right|$ is bounded,
it easy to extend to complex-valued $V$; Its proof is almost same
as that of the real-valued cases in \cite{Gun10}.

Set $M=\rG$, and consider the Brownian motion $B$ on the Riemannian
manifold $\rG$ in the time interval $[0,\infty)$, where the distribution
of the starting point is uniform on $\rG$, i.e. equals the Haar measure
$\d g$ on $\rG$. Let $\mu^{1}$ be a probability measure on $C([0,\infty),\rG)$
which represents such Brownian motion (i.e. a Wiener measure uniform
on $\rG$). Then the above theorem is restated as

\begin{shaded}%
\begin{prop}
\label{thm:<f2|e-tHf1>=00003Dint[]dmu}\displabel{thm:<f2|e{-}tHf1>=00003Dint{[}{]}dmu}Let $V\in C(\rG,\C)$. For
any $f_{1},f_{2}\in L^{2}(\rG)$ and $t\ge0$,
\[
\bigl\langle f_{2}|e^{-tH(\alpha,V)}f_{1}\bigr\rangle=\int_{C([0,\infty),\rG)}\left[e^{\cI_{\alpha,V}}\ol{f_{2}(B_{0})}f_{1}(B_{t})\right]\d\mu^{1}(B).\ 
\]
\end{prop}
\end{shaded}

For $\ss>0$, let $B_{t}^{\ss}:=B_{\ss t}$, and define the probability
measure $\mu^{\ss}$ on $C([0,\infty),\rG)$ by
\[
\d\symb{\mu^{\ss}}\index{${mun}$@${\mu^{\ss}}$}\hypertarget{ref.84}{}(B_{t}^{\ss}):=\d\mu^{1}(B_{t}),
\]
Recall $\symb{\alpha}\index{${alph}$@${\alpha}$}\hypertarget{ref.85}{}:=-(\lambda+\rho)$ and $\symb{c_{\lambda}}\index{${clam}$@${c_{\lambda}}$}\hypertarget{ref.86}{}=\inf{\rm spec}\Delta^{-(\lambda+\rho)}=\left\langle 2\lambda+\rho|\rho\right\rangle _{\fg_{\C}^{*}}.$

\begin{shaded}%
\begin{thm}
\label{thm:<f2|e-tSf1>=00003Deint}\displabel{thm:<f2|e{-}tSf1>=00003Deint}Let
\[
\symb{S_{\ss}(h)}\index{${Snum(h)}$@${S_{\ss}(h)}$}\hypertarget{ref.87}{}:=\frac{1}{2}\ss\ul{\Delta}^{-(\lambda+\rho)}+\im\fextend h.
\]
Then for $h\in C^{\infty}(\rG\cdot\Hwvproj,\R)$, $f_{1},f_{2}\in\cH_{\lambda}(\rG)=\ker\ul{\Delta}^{-(\lambda+\rho)}$
and $t\ge0$,%
\begin{equation}
\bigl\langle f_{2}|\ e^{-tS_{\ss}(h)}f_{1}\bigr\rangle=e^{\frac{1}{2}\ss tc_{\lambda}}\int_{C([0,\infty),\rG)}\left[e^{\cI_{t}(h)}\ol{f_{2}(B_{0})}f_{1}(B_{t})\right]\d\mu^{\ss}(B).\ \label{eq:<f1|exp(-tS)f2>}\displabel{eq:<f1|exp({-}tS)f2>}
\end{equation}
where
\[
\symb{\cI_{t}(h)}\index{${Ih0t}$@${\cI_{t}(h)}$}\hypertarget{ref.88}{}:=\int_{0}^{t}\alpha(\du B_{s})-\im\int_{0}^{t}\fextend h(B_{s})\d s.
\]
\end{thm}
\end{shaded}

\begin{proof}
Let $V:=-\frac{1}{2}c_{\lambda}+\im\frac{1}{\ss}\fextend h$. Then
we see
\[
S_{\ss}(h)=\ss\left(\frac{1}{2}\Delta^{\alpha}+V\right)=\ss H(\alpha,V).
\]
Let $\mathsf{W}:=C([0,\infty),\rG)$. Then by Prop. \ref{thm:<f2|e-tHf1>=00003Dint[]dmu},
we have
\begin{align*}
 & \bigl\langle f_{2}|\ e^{-tS_{\ss}(h)}f_{1}\bigr\rangle=\left\langle f_{2}|e^{-t\ss H(\alpha,V)}f_{1}\right\rangle \ \ \\
 & \quad=\int_{\mathsf{W}}\left[\exp\left(\int_{0}^{\ss t}\alpha(\du B_{s})-\int_{0}^{\ss t}V(B_{s})\d s\right)\ol{f_{2}(B_{0})}f_{1}(B_{\ss t})\right]\d\mu^{1}(B)\\
 & \quad=\int_{\mathsf{W}}\left[\exp\left(\int_{0}^{\ss t}\alpha(\du B_{s})+\frac{1}{2}\ss tc_{\lambda}-\im\frac{1}{\ss}\int_{0}^{\ss t}\fextend h(B_{s})\d s\right)\ol{f_{2}(B_{0})}f_{1}(B_{\ss t})\right]\d\mu^{1}(B)\\
 & \quad=\int_{\mathsf{W}}\left[\exp\left(\int_{0}^{t}\alpha(\du B_{s})+\frac{1}{2}\ss tc_{\lambda}-\im\int_{0}^{t}\fextend h(B_{s})\d s\right)\ol{f_{2}(B_{0})}f_{1}(B_{t})\right]\d\mu^{\ss}(B)\\
 & \quad=e^{\frac{1}{2}\ss tc_{\lambda}}\int_{\mathsf{W}}\left[\exp\left(\int_{0}^{t}\alpha(\du B_{s})-\im\int_{0}^{t}\fextend h(B_{s})\d s\right)\ol{f_{2}(B_{0})}f_{1}(B_{t})\right]\d\mu^{\ss}(B).\ \ 
\end{align*}

\end{proof}
Fix an arbitrary $f\in\cH_{\lambda}(\rG)$ with $\left\Vert f\right\Vert =1$.
If we set $h\equiv0$ in (\ref{eq:<f1|exp(-tS)f2>}), since $e^{-tS_{\ss}(0)}f=f$,
we see that the `normalization factor' $e^{-\frac{1}{2}\ss tc_{\lambda}}$
can be included in the integral measure:

\[
\symb{Z_{\lambda,t,\ss}}\index{${Zlambt}$@${Z_{\lambda,t,\ss}}$}\hypertarget{ref.89}{}:=e^{-\frac{1}{2}\ss tc_{\lambda}}=\int_{C([0,\infty),\rG)}\left[e^{\cI_{t}(0)}\ol{f(B_{0})}f(B_{t})\right]\d\mu^{\ss}(B).
\]

\begin{shaded}%
\begin{cor}
[Brownian path integral]\label{thm:main-Brown2}\displabel{thm:main{-}Brown2}For $h\in C^{\infty}(\rG\cdot\Hwvproj,\R)$,
$f_{1},f_{2}\in\cH_{\lambda}(\rG)$ and $t\ge0$,%
{} we have
\[
\bigl\langle f_{2}|e^{\im t\cQ(h)}f_{1}\bigr\rangle=\lim_{\ss\to\infty}\int_{C([0,\infty),\rG)}\left[e^{\cI_{t}(h)}\ol{f_{2}(B_{0})}f_{1}(B_{t})\right]\frac{\d\mu^{\ss}(B)}{Z_{\lambda,t,\ss}}.\ 
\]
\end{cor}
\end{shaded}
\begin{proof}
Directly follows from the asymptotic representation theorem \ref{thm:sibori-G}
and Theorem \ref{thm:<f2|e-tSf1>=00003Deint}.%

\end{proof}

\section{Rough path theory}

\label{sec:roughPath}\displabel{sec:roughPath}

In the study of stochastic processes, the It\^o Calculus, based on
martingale theory, has been the most effective tool for many years.
But a few alternative (or additional) approaches are known; e.g. the
Malliavin Calculus, and \emph{rough paths theory} which we use in
this paper. %
{} Among other things, rough path theories have made a considerable
progress on the problem of the (piecewise) smooth approximations of
stochastic processes. This problem is an old but also up-to-date one,
since it is related to the problem of \emph{renormalization} occurring
mainly in quantum physics. (Another rigorous approach to renormalization
is lattice field theory.) When one considers the problem to approximate
a martingale by a sequence of other martingales, conventional martingale
theory will suffice. However, since a (piecewise) smooth process is
not a martingale, it is difficult to deal with smooth approximations
in martingale theory (see the complicated analysis in \cite{I-W89}).
One will find in next section that the theory of geometric rough paths
is the best approach to such problems.

Rough path theory was originated by Lyons \cite{Lyo98}, and has been
extensively developed into several approaches, including the large-scale
theories such as the %
theory of Gubinelli--Imkeller--Perkowski \cite{GIP15}, and %
that of Hairer \cite{Hai14}. So it seems impossible to give a brief
overview of rough path theories. %
(Different approaches use different definitions of the fundamental
notions such as `rough integral' and `rough differential equation'.)
Instead we refer to a single approach of Friz--Victoir book \cite{FV10b}.
However, since this 650-pages book is not easily accessible for everyone,
we will summarize their approach here for the convenience of readers.
See also Baudoin's lecture note \cite{Bau13}, which is more concise
and accessible.

Let $\VECSP\cong\R^{d}$ be a vector space with the usual norm, and
$\Tensor(\VECSP)$ be the tensor algebra over $\VECSP$, i.e.,
\[
\symb{\Tensor(\VECSP)}\index{${T(V)}$@${\Tensor(\VECSP)}$}\hypertarget{ref.90}{}:=\bigoplus_{k=0}^{\infty}\Tensor^{k}(\VECSP),\quad\symb{\Tensor^{k}(\VECSP)}\index{${Tk(V)}$@${\Tensor^{k}(\VECSP)}$}\hypertarget{ref.91}{}:=\VECSP^{\otimes k}.
\]
Let 
\[
\symb{\Tensor^{\le N}(\VECSP)}\index{${Tn(V)}$@${\Tensor^{\le N}(\VECSP)}$}\hypertarget{ref.92}{}:=\bigoplus_{k=0}^{N}\Tensor^{k}(\VECSP),
\]
and $\symb{\projection_{k}}\index{${prN}$@${\projection_{k}}$}\hypertarget{ref.93}{}$ and $\symb{\projection_{\le N}}\index{${prN}$@${\projection_{\le N}}$}\hypertarget{ref.94}{}$
denote the projection from $\Tensor(\VECSP)$ onto $\Tensor^{k}(\VECSP)$
and $\Tensor^{\le N}(\VECSP)$, respectively. We make $\Tensor^{\le N}(\VECSP)$
into an algebra with the product defined by
\[
xy:=\projection_{\le N}(x\otimes y)\in\Tensor^{\le N}(\VECSP),\qquad x,y\in\Tensor^{\le N}(\VECSP)
\]
$\Tensor^{\le N}(\VECSP)$ is called the \term{truncated tensor algebra}\index{zzz@truncated tensor algebra}\hypertarget{truncated+tensor+algebra}{}.
$\Tensor^{\le N}(\VECSP)$ is also a Lie algebra with the Lie bracket
$[x,y]=xy-yx$.%
{} Define $\symb{\fg_{N}(\VECSP)}\index{${gN}$@${\fg_{N}(\VECSP)}$}\hypertarget{ref.95}{}\subset\Tensor^{\le N}(\VECSP)$
as the Lie subalgebra of $\Tensor^{\le N}(\VECSP)$ generated by $\VECSP=\Tensor^{1}(\VECSP)\subset\Tensor^{\le N}(\VECSP)$.
{} Define the Lie group $\symb{\bbG_{N}(\VECSP)}\index{${GN(V)}$@${\bbG_{N}(\VECSP)}$}\hypertarget{ref.96}{}\subset\Tensor^{\le N}(\VECSP)$
by
\[
\symb{\bbG_{N}(\VECSP)}\index{${GN(V)}$@${\bbG_{N}(\VECSP)}$}\hypertarget{ref.97}{}:=\exp\left(\fg_{N}(\VECSP)\right)=\left\{ \sum_{n=0}^{N}\frac{x^{n}}{n!}:\ x\in\fg_{N}(\VECSP)\right\} .
\]
$\bbG_{N}(\VECSP)$ is called the \term{free nilpotent group}\index{zzz@free nilpotent group}\hypertarget{free+nilpotent+group}{} of
step $N$. We see

\[
\bbG_{2}(\VECSP)=\left\{ 1+v+\frac{1}{2}v^{2}+A:\ v\in\VECSP,\ A\in{\rm Anti}(\VECSP^{\otimes2})\right\} 
\]
where ${\rm Anti}(\VECSP^{\otimes2})$ is the subspace of $\VECSP^{\otimes2}$
spanned by $\left\{ u\otimes v-v\otimes u:\ u,v\in\VECSP\right\} $.

Let $\symb{C^{1\hvar}([0,T],\VECSP)}\index{${C1-var}$@${C^{1\hvar}([0,T],\VECSP)}$}\hypertarget{ref.98}{}$ denote the subspace
of $C([0,T],\VECSP)$ consisting of the functions of bounded variation.
Let $x\in C^{1\hvar}([0,T],\VECSP)$. For $n=0,1,...$ and $0\le s<t$,
define $x_{s,t}^{\{n\}}\in\VECSP^{\otimes n}$ by
\begin{equation}
\symb{x_{s,t}^{\{0\}}}\index{${x\{0\}}$@${x_{s,t}^{\{0\}}}$}\hypertarget{ref.99}{}:=1,\quad\symb{x_{s,t}^{\{n\}}}\index{${x\{n\}}$@${x_{s,t}^{\{n\}}}$}\hypertarget{ref.100}{}:=\int_{s<u_{1}<\cdots<u_{n}<t}\d x_{u_{1}}\otimes\cdots\otimes\d x_{u_{n}},\quad n\ge1.\label{eq:def:xs,t{n}}\displabel{eq:def:xs,t\{n\}}
\end{equation}
where the integral is of the sense of Riemann--Stieltjes. We see that
$x_{s,t}^{\{n\}}$, $n=1,2$ is explicitly written as
\[
\symb{x_{s,t}^{\{1\}}}\index{${x\{1\}}$@${x_{s,t}^{\{1\}}}$}\hypertarget{ref.101}{}=\symb{x_{s,t}}\index{${xst}$@${x_{s,t}}$}\hypertarget{ref.102}{}:=x_{t}-x_{s},\quad\symb{x_{s,t}^{\{2\}}}\index{${x\{2\}}$@${x_{s,t}^{\{2\}}}$}\hypertarget{ref.103}{}=\int_{s}^{t}\left(x_{r}-x_{s}\right)\otimes\d x_{r}.\ 
\]
The \term{step-$N$ signature}\index{zzz@step-$N$ signature}\hypertarget{step-_24N_24+signature}{} of $x$ is given by 
\[
\symb{S_{N}(x)_{s,t}}\index{${SN()}$@${S_{N}(x)_{s,t}}$}\hypertarget{ref.104}{}\equiv\symb{x_{s,t}^{\{\le N\}}}\index{${x[<N]}$@${x_{s,t}^{\{\le N\}}}$}\hypertarget{ref.105}{}:=\bigoplus_{k=0}^{N}x_{s,t}^{\{k\}}\in\Tensor^{\le N}(\VECSP),\quad t\in[0,T].
\]
In fact $S_{N}(x)=x^{\{\le N\}}$ is a path on the free nilpotent
group $\bbG_{N}(\VECSP)\subsetneq\Tensor^{\le N}(\VECSP)$; Precisely,
it is shown that
\begin{align*}
\bbG_{N}(\VECSP) & =\left\{ S_{N}(x)_{s,t}:\ x\in C^{1\hvar}([0,T],\VECSP),\ 0\le s<t\le T\right\} \ \\
 & =\left\{ S_{N}(x)_{0,1}:\ x\in C^{1\hvar}([0,1],\VECSP)\right\} .
\end{align*}

Then we have the following fundamental algebraic relation:

\begin{shaded}%
\begin{thm}
[Chen's relation]\label{thm:Chen'sTheorem-FV}\displabel{thm:Chen{'}sTheorem{-}FV}%
{} Given $x\in C^{1\hvar}\big([0,T],\VECSP\big)$ and $0\le s<t<u\le T$
we have %
\[
S_{N}(x)_{s,u}=S_{N}(x)_{s,t}S_{N}(x)_{t,u},
\]
where the rhs is the product in the free nilpotent group $\bbG_{N}(\VECSP)$
($=$the product in the truncated tensor algebra $\Tensor^{\le N}(\VECSP)$).\end{thm}
\end{shaded}%
For any $x\in C^{1\hvar}([0,T],\VECSP)$ and $0\le t_{1}<t_{2}\le T$,
any path segment%

\[
S_{N}(x)_{0,\bullet}\big|_{[t_{1},t_{2}]}:\,[t_{1},t_{2}]\ni t\longmapsto S_{N}(x)_{0,t}\in\bbG_{N}(\VECSP),
\]
as well as its reparametrizations, is said to be \term{horizontal}\index{zzz@horizontal}\hypertarget{horizontal}{}.
{} It is shown that any two points of $\bbG_{N}(\VECSP)$ can be connected
by a horizontal path, and hence we can define a ``geodesic distance''
$d_{{\rm CC}}$ of two points $g,h\in\bbG_{N}(\VECSP)$ as follows:
\begin{align}
 & \symb{d_{{\rm CC}}(g,h)}\index{${dCC}$@${d_{{\rm CC}}(g,h)}$}\hypertarget{ref.106}{}\notag\\
 & \quad:=\inf\Bigl\{{\rm length}\left(x|_{[t_{1},t_{2}]}\right):\ 0\le t_{1}<t_{2}\le T,\ \notag\\
 & \qquad\qquad x\in C^{1\hvar}([0,T],\VECSP),\ x_{0,t_{1}}^{\{\le N\}}=g,\ x_{0,t_{2}}^{\{\le N\}}=h\Bigr\}\\
 & \quad=\inf\left\{ {\rm length}\left(x|_{[0,1]}\right):\, x\in C^{1\hvar}([0,1],\VECSP),\ x_{0,1}^{\{\le N\}}=g^{-1}h\right\} .\ \label{eq:def:Carnot-CaratheodoryD}\displabel{eq:def:Carnot{-}CaratheodoryD}
\end{align}
where the length of the path $x|_{[t_{1},t_{2}]}$ is usually defined
by the metric on $\VECSP$. In fact $d_{{\rm CC}}$ turns out to be
a metric on $\bbG_{N}(\VECSP)$, and is called the \term{Carnot--Carath\'eodory metric}\index{zzz@Carnot--Carath\'eodory metric}\hypertarget{Carnot--Carath_5C_27eodory+metric}{}.

If $s\le t$ we will denote by $\symb{\dissect[s,t]}\index{${Delta[]}$@${\dissect[s,t]}$}\hypertarget{ref.107}{}$, the
set of subdivisions of the interval $[s,t]$, that is $\Pi\in\dissect[s,t]$
can be written
\[
\Pi=\left\{ s=t_{0}<t_{1}<\cdots<t_{n}=t\right\} .
\]

\begin{defn}
{} Let $\cG$ be a group with the unit $1_{\cG}\in\cG$, and a left
invariant metric $d$ on $\cG$. %
For a path ${\bf x}:[0,T]\to\cG$, let $\symb{{\bf x}_{s,t}}\index{${xs,t}$@${{\bf x}_{s,t}}$}\hypertarget{ref.108}{}:={\bf x}_{s}^{-1}{\bf x}_{t}\in\cG.$
For ${\bf x},{\bf y}:[0,T]\to\cG$ and $p>0$, the \term{$p$-variation distance}\index{zzz@$p$-variation distance}\hypertarget{_24p_24-variation+distance}{}
(semi-metric) between ${\bf x}$ and ${\bf y}$ is defined by
\begin{equation}
\symb{d_{\cG,p\hvar;[0,T]}({\bf x},{\bf y})}\index{${dG,p-var;[]}$@${d_{\cG,p\hvar;[0,T]}({\bf x},{\bf y})}$}\hypertarget{ref.109}{}:=\Bigl[\sup_{\{t_{i}\}\in\dissect[0,T]}\sum_{i}d\left({\bf x}_{t_{i},t_{i+1}},{\bf y}_{t_{i},t_{i+1}}\right)^{p}\Bigr]^{1/p}\label{eq:def:d_{p-var}-general}\displabel{eq:def:d\_\{p{-}var\}{-}general}
\end{equation}
A path ${\bf x}:[0,T]\to\cG$ is said to be of \term{finite $p$-variation}\index{zzz@finite $p$-variation}\hypertarget{finite+_24p_24-variation}{}
if $d_{\cG,p\hvar;[0,T]}(\boldsymbol{1}_{\cG},{\bf x})<\infty$, where
$\boldsymbol{1}_{\cG}$ is the constant path with value $1_{\cG}$.
The space of the paths of finite $p$-variation is denoted by $\symb{C^{p\hvar}([0,T],\cG)}\index{${Cp-var(cG)}$@${C^{p\hvar}([0,T],\cG)}$}\hypertarget{ref.110}{}$.
The \term{$p$-variation metric}\index{zzz@$p$-variation metric}\hypertarget{_24p_24-variation+metric}{} on $C^{p\hvar}([0,T],\cG)$ is given
by 
\[
\symb{\tilde{d}_{\cG,p\hvar;[0,T]}({\bf x},{\bf y})}\index{${dG,p-bartil}$@${\tilde{d}_{\cG,p\hvar;[0,T]}({\bf x},{\bf y})}$}\hypertarget{ref.111}{}:=d({\bf x}_{0},{\bf y}_{0})+d_{\cG,p\hvar;[0,T]}({\bf x},{\bf y}),
\]
which determines a topology on $C^{p\hvar}([0,T],\cG)$. Let 
\[
\symb{C_{0}^{p\hvar}([0,T],\cG)}\index{${Cp-var(cG)}$@${C_{0}^{p\hvar}([0,T],\cG)}$}\hypertarget{ref.112}{}:=\left\{ x\in C^{p\hvar}([0,T],\cG):\ x_{0}=1_{\cG}\right\} .\ 
\]

\end{defn}
{}If $\cG$ is the additive group $\VECSP\cong\R^{d}$, we see
\[
d_{\VECSP,p\hvar;[0,T]}({\bf x},{\bf y})=\left\Vert {\bf y}-{\bf x}\right\Vert _{p\hvar;[0,T]}
\]
where $\left\Vert \bullet\right\Vert _{p\hvar;[0,T]}$ denotes the
$p$-\term{variation seminorm}\index{zzz@variation seminorm}\hypertarget{variation+seminorm}{} defined by
\[
\symb{\left\Vert {\bf x}\right\Vert _{p\hvar;[0,T]}}\index{${||x||p-var}$@${\left\Vert {\bf x}\right\Vert _{p\hvar;[0,T]}}$}\hypertarget{ref.113}{}:=\Bigl(\sup_{\{t_{i}\}\in\dissect[0,T]}\sum_{k}\left\Vert {\bf x}_{t_{k+1}}-{\bf x}_{t_{k}}\right\Vert ^{p}\Bigr)^{1/p}.
\]
Set $\cG=\bbG_{N}(\VECSP)$ with the Carnot--Carath\'eodory metric
$d_{{\rm CC}}$, and consider the path space $C^{p\hvar}([0,T],\bbG_{N}(\VECSP))$,
called the space of \term{weak geometric $p$-rough paths}\index{zzz@weak geometric $p$-rough paths}\hypertarget{weak+geometric+_24p_24-rough+paths}{}. %
For $p\ge1$, this is a complete, non-separable metric space \cite[p.175 Theorem 8.13]{FV10b}.

Define $\symb{\GEOMR_{[0,T],0}^{p}(\VECSP)}\index{${GR0()}$@${\GEOMR_{[0,T],0}^{p}(\VECSP)}$}\hypertarget{ref.114}{}$ (which is denoted
by $\symb{C_{0}^{0,p\hvar}\big([0,T],G^{\left\lfloor p\right\rfloor }\big(\VECSP\big)\big)}\index{${C0,p-var(GN)}$@${C_{0}^{0,p\hvar}\big([0,T],G^{\left\lfloor p\right\rfloor }\big(\VECSP\big)\big)}$}\hypertarget{ref.115}{}$
in \cite{FV10b}, $\symb{\bOmega\bG_{0}^{p}([0,T],\VECSP)}\index{${OmegaGp}$@${\bOmega\bG_{0}^{p}([0,T],\VECSP)}$}\hypertarget{ref.116}{}$
in \cite{Bau13}) to be the set of continuous paths ${\bf x}:[0,T]\to\bbG_{\left\lfloor p\right\rfloor }\big(\VECSP\big)$
for which there exists a sequence $x_{n}\in C^{\infty}([0,T],\VECSP)$
such that%

\[
\lim_{n\to\infty}S_{\left\lfloor p\right\rfloor }(x_{n})={\bf x}\quad\text{in }\left(C_{0}^{p\hvar}\big([0,T],\bbG_{\left\lfloor p\right\rfloor }\big(\VECSP\big)\big),d_{p\hvar;[0,T]}\right).
\]
{} Let
\[
\symb{\GEOMR_{[0,T]}^{p}(\VECSP)}\index{${OmegaGp}$@${\GEOMR_{[0,T]}^{p}(\VECSP)}$}\hypertarget{ref.117}{}:=\left\{ {\bf x}:[0,T]\to\bbG_{\left\lfloor p\right\rfloor }(\VECSP):\ {\bf x}_{0,\bullet}\in\GEOMR_{[0,T],0}^{p}(\VECSP)\right\} .
\]
(Recall ${\bf x}_{0,t}:={\bf x}_{0}^{-1}{\bf x}_{t}$.) In other words,
$\GEOMR_{[0,T]}^{p}(\VECSP)$ is the $\tilde{d}_{p\hvar;[0,T]}$-closure
of $C^{\infty}\big([0,T],\bbG_{\left\lfloor p\right\rfloor }\big(\VECSP\big)\big)$.
An element of $\GEOMR_{[0,T]}^{p}(\VECSP)$ is called a \term{geometric $p$-rough path}\index{zzz@geometric $p$-rough path}\hypertarget{geometric+_24p_24-rough+path}{}.
For $p>1$, a geometric $p$-rough path is characterized as a path
$[0,T]\to\bbG_{\left\lfloor p\right\rfloor }\big(\VECSP\big)$ which
is \term{absolutely continuous of order $p$}\index{zzz@absolutely continuous of order $p$}\hypertarget{absolutely+continuous+of+order+_24p_24}{}, or \term{$p$-absolutely continuous}\index{zzz@$p$-absolutely continuous}\hypertarget{_24p_24-absolutely+continuous}{}
(in the sense of Wiener--Young--Love) (\cite[pp.96,180]{FV10b}, see
also \cite{DN99,ABM13}). It is shown that $\left(\GEOMR_{[0,T]}^{p}(\VECSP),\tilde{d}_{p\hvar;[0,T]}\right)$
is a complete separable metric space \cite[p.180, Proposition 8.25]{FV10b}. 

Remark: Since $\bbG_{N}(\VECSP)$ is a subset of the normed linear
space $\Tensor^{\le N}(\VECSP)$, $\bbG_{N}(\VECSP)$ also has a metric
$d_{\Tensor}$ of $\Tensor^{\le N}(\VECSP)$, different from $d_{{\rm CC}}$.
However, $d_{\Tensor}$ is not a left invariant metric on $\bbG_{N}(\VECSP)$,
and hence we cannot replace $d_{{\rm CC}}$ with $d_{\Tensor}$.%

Let $\varphi:\R^{d_{1}}\to L(\R^{d_{2}},\R^{d_{3}})$, where $L(\R^{d_{2}},\R^{d_{3}})$
is the space of linear maps $\R^{d_{2}}\to\R^{d_{3}}$. If $\R^{d_{1}}\ni x\mapsto\varphi(x)e_{k}\in\R^{d_{3}}$
is $\gamma$-Lipschitz for all $k=1,...,d_{2}$, where $(e_{k})$
is the standard basis of $\R^{d_{2}}$, we write $\varphi\in\symb{\Lip^{\gamma}(\R^{d_{1}},L(\R^{d_{2}},\R^{d_{3}}))}\index{${Lipgamma}$@${\Lip^{\gamma}(\R^{d_{1}},L(\R^{d_{2}},\R^{d_{3}}))}$}\hypertarget{ref.118}{}$.

Let $\gamma>p$, $V\in\Lip^{\gamma}(\R^{e},L(\R^{d},\R^{e}))$ and
$x\in C^{1\hvar}([0,T],\R^{d})$. Then there exists a unique solution
$y\in C^{1\hvar}([0,T],\R^{e})$ of the ordinary differential equation
(ODE) 
\[
\d y(t)=V(y(t))\d x(t),\quad y(0)=y_{0}\in\R^{e}.
\]
Thus we define the map
\begin{align*}
\R^{e}\times C^{1\hvar}([0,T],\R^{d}) & \longrightarrow C^{1\hvar}([0,T],\R^{e})\\
(y_{0},x) & \longmapsto\pi_{(V)}(y_{0},x)
\end{align*}
by $\symb{\pi_{(V)}(y_{0},x)}\index{${pi(V)}$@${\pi_{(V)}(y_{0},x)}$}\hypertarget{ref.119}{}:=y$.

We assume moreover $V\in C^{\infty}$ here; In the next section it
will suffice to consider only the case where $V$ is smooth.

\begin{shaded}%
\begin{thm}
\label{thm:RDEsol}\displabel{thm:RDEsol} Let $p\ge1$ and $N=\left\lfloor p\right\rfloor $.
Let $x_{n}\in C^{\infty}([0,T],\R^{d})$, ${\bf x}_{n}:=S_{\left\lfloor p\right\rfloor }(x_{n})$
for each $n\in\N$, and ${\bf x}\in\GEOMR_{[0,T]}^{p}(\R^{d})$, and
assume
\[
\lim_{n\to\infty}{\bf x}_{n}={\bf x}\quad\text{in }\left(\GEOMR_{[0,T]}^{p}(\R^{d}),\,\tilde{d}_{p\hvar;[0,T]}\right).
\]
Then the limit
\[
\symb{\pi_{(V)}(y_{0},{\bf x})}\index{${pi(V)}$@${\pi_{(V)}(y_{0},{\bf x})}$}\hypertarget{ref.120}{}:=\lim_{n\to\infty}\pi_{(V)}(y_{0},x_{n})
\]
converges in $\left(C^{p\hvar}([0,T],\R^{e}),\,\tilde{d}_{p\hvar;[0,T]}\right)$.
Furthermore, for each ${\bf y}_{0}\in\bbG_{\left\lfloor p\right\rfloor }(\R^{e})$
with $\projection_{1}({\bf y}_{0})=y_{0}$, the limit 
\[
\symb{\bpi_{(V)}({\bf y}_{0},{\bf x})}\index{${pi(V)}$@${\bpi_{(V)}({\bf y}_{0},{\bf x})}$}\hypertarget{ref.121}{}:=\lim_{n\to\infty}{\bf y}_{0}S_{\left\lfloor p\right\rfloor }\left(\pi_{(V)}(y_{0},x_{n})\right)
\]
converges in $\left(\GEOMR_{[0,T]}^{p}(\R^{e}),\,\tilde{d}_{p\hvar;[0,T]}\right)$,
and satisfies
\[
\pi_{(V)}(y_{0},{\bf x})=\projection_{1}\left(\bpi_{(V)}({\bf y}_{0},{\bf x})\right).
\]
These definitions of $\pi_{(V)}(y_{0},{\bf x})$ and $\bpi_{(V)}({\bf y}_{0},{\bf x})$
do not depend on the choice of the approximating sequence ${\bf x}_{n}$.

(Make sure to distinguish between $\pi_{(V)}$ and bold letter $\bpi_{(V)}$.) \end{thm}
\end{shaded}

We call $y:=\pi_{(V)}(y_{0},{\bf x})$ (resp. ${\bf y}:=\bpi_{(V)}({\bf y}_{0},{\bf x})$)
the \term{solution of the rough differential equation (RDE solution)}\index{zzz@solution of the rough differential equation (RDE solution)}\hypertarget{solution+of+the+rough+differential+equation+_28RDE+solution_29}{}
(resp. the \term{full RDE solution}\index{zzz@full RDE solution}\hypertarget{full+RDE+solution}{}) of
\begin{equation}
\d y(t)=V(y(t))\d x(t)\quad\text{resp.}\quad\d{\bf y}(t)=V({\bf y}(t))\d{\bf x}(t),\label{eq:RDEdiff}\displabel{eq:RDEdiff}
\end{equation}
with $y(0)=y_{0}\in\R^{e}.$ (resp. ${\bf y}(0)={\bf y}_{0}$), and
call the map $\pi_{(V)}$ (resp. $\bpi_{(V)}$) the \term{It\^o--Lyons map}\index{zzz@It\^o--Lyons map}\hypertarget{It_5C_5Eo--Lyons+map}{}
(resp. \term{full It\^o--Lyons map}\index{zzz@full It\^o--Lyons map}\hypertarget{full+It_5C_5Eo--Lyons+map}{}). 

The above definition of (full) RDE solution is slightly modified version
of \cite[p.70]{Bau13}, which is slightly different from that of \cite[p.224]{FV10b},

The full It\^o--Lyons map is characterized as the extension of the
ODE solution map $\pi_{(V)}$ which satisfies the following continuity:

\begin{shaded}%
\begin{thm}
Let $d_{1}:=\tilde{d}_{p\hvar;[0,T]}$ and $d_{2}:=d_{\infty;[0,T]}$,
where 
\[
\symb{d_{\infty;[0,T]}({\bf x},{\bf y})}\index{${d_{\infty;[0,T]}({\bf x},{\bf y})}$@${d_{\infty;[0,T]}({\bf x},{\bf y})}$}\hypertarget{ref.122}{}:=\sup_{t\in[0,T]}d_{{\rm CC}}({\bf x}_{t},{\bf y}_{t}).
\]
Then the full It\^o--Lyons map
\begin{align*}
\bbG_{\left\lfloor p\right\rfloor }(\R^{e})\times\Big(\GEOMR_{[0,T]}^{p}(\R^{d}),d_{k}\Big) & \to\Big(\GEOMR_{[0,T]}^{p}(\R^{e}),d_{k}\Big)\\
({\bf y}_{0},{\bf x}) & \mapsto\boldsymbol{\pi}_{(V)}({\bf y}_{0};{\bf x})
\end{align*}
is continuous for $k=1,2$. In fact, these are uniformly continuous
on each $\tilde{d}_{p\hvar}$-bounded sets.\end{thm}
\end{shaded}

Let $\VECSP_{1}=\R^{d},\ \VECSP_{2}=\R^{e}$, and $\varphi\in\Lip^{\gamma-1}(\VECSP_{1},L(\VECSP_{1},\VECSP_{2}))\cap C^{\infty}$.
Define $\Phi:\VECSP_{1}\oplus\VECSP_{2}\to L(\VECSP_{1},\VECSP_{1}\oplus\VECSP_{2})$
by%
\[
\Phi(x\oplus y)x':=x'\oplus\varphi(x)x',\quad x,x'\in\VECSP_{1},\ y\in\VECSP_{2}.
\]
Then we easily see
\[
{\rm proj}_{\VECSP_{2}}\pi_{(\Phi)}(0,x)=\int_{0}^{\bullet}\varphi(x(s))\d x(s),\quad x\in C^{1\hvar}([0,T],\VECSP_{1})
\]
where ${\rm proj}_{\VECSP_{2}}$ is the projection from $\VECSP_{1}\oplus\VECSP_{2}$
onto $\VECSP_{2}$. Thus the Riemann--Stieltjes line integral $\int\varphi(x)dx$
can be expressed by the ODE solution map $\pi_{(\Phi)}$. Similarly
we define the \term{rough line integral}\index{zzz@rough line integral}\hypertarget{rough+line+integral}{} $\symb{\bUpsilin_{(\varphi)}({\bf x})}\index{${I(phi)}$@${\bUpsilin_{(\varphi)}({\bf x})}$}\hypertarget{ref.123}{}\equiv\int_{0}^{\bullet}\varphi({\bf x})\d{\bf x}$
for ${\bf x}\in\GEOMR_{[0,T]}^{p}(\VECSP_{1})$ to be a map

\[
\bUpsilin_{(\varphi)}:\GEOMR_{[0,T]}^{p}(\VECSP_{1})\to\GEOMR_{[0,T]}^{p}(\VECSP_{2})
\]
defined by

\begin{equation}
\symb{\bUpsilin_{(\varphi)}({\bf x})}\index{${I(phi)}$@${\bUpsilin_{(\varphi)}({\bf x})}$}\hypertarget{ref.124}{}\equiv\int_{0}^{\bullet}\varphi({\bf x}(s))\d{\bf x}(s):={\rm proj}_{\Tensor^{\le N}(\VECSP_{2})}\bpi_{(\Phi)}\left(0,{\bf x}\right).\label{eq:def:roughInt}\displabel{eq:def:roughInt}
\end{equation}

\section{Brownian motion as rough path}

\label{sec:BrownAsRP}\displabel{sec:BrownAsRP}

{} Let $B_{t}$ be a Brownian motion (or more generally a semimartingale)
on $\VECSP=\R^{d}$. Then $B\notin C^{1\hvar}([0,T],\VECSP)$ a.s.,
and hence the step-$N$ signature $S_{N}(B)_{s,t}\equiv B_{s,t}^{\{\le N\}}$
is not defined by the Riemann--Stieltjes integral of (\ref{eq:def:xs,t{n}})
when $N\ge2$. However we find that if we set 
\[
\bB_{s,t}:=1\oplus B_{s,t}\oplus\int_{s}^{t}B\otimes\du B,
\]
where $\du B$ denotes Stratonovich integration, then $\bB_{s,t}\in\bbG_{2}\big(\VECSP\big)$.
In fact it is shown that $\bB$ is a geometric $p$-rough path for
$2<p<3$, i.e., $\bB:=\bB_{0,\bullet}\in\GEOMR_{[0,T]}^{p}(\VECSP)$,
almost surely. Thus a Brownian motion $B$ in $[0,T]$ can be identified
with a $\GEOMR_{[0,T]}^{p}(\VECSP)$-valued random variable $\bB$,
called the \term{enhanced Brownian motion}\index{zzz@enhanced Brownian motion}\hypertarget{enhanced+Brownian+motion}{}. Moreover, the solution
of the stochastic differential equation%
{} $\d Y_{t}=V(Y_{t})\du B_{t}$ can be identified with the solution
of the RDE $\d Y_{t}=V(Y_{t})\d B_{s}$; Precisely,

\begin{shaded}%
\begin{thm}
{\rm \cite[p.510 Theorem 17.3]{FV10b}} Let $p,\gamma$ be such
that $2<p<\gamma.$ Let $V\in\Lip^{\gamma}(\R^{e},L(\R^{d},\R^{e}))$,
$y_{0}\in\R^{e}$ and $B$ be an $\R^{d}$-valued semimartingale,
enhanced to $\bB=\bB(\omega)\in\GEOMR_{[0,T]}^{p}(\R^{d})$ almost
surely. Then the (for a.e. $\omega$ well-defined) RDE solution
\[
Y(\omega)=\pi_{(V)}(y_{0};\bB(\omega)),
\]
solves the Stratonovich SDE
\begin{equation}
\d Y=V(Y)\du B,\quad Y(0)=y_{0}.\label{eq:StratDE-diff}\displabel{eq:StratDE{-}diff}
\end{equation}
\end{thm}
\end{shaded}

Note that the definitions of RDE solution (and rough integral) do
not refer to any probability measure, i.e., they are deterministically
defined. Hence this viewpoint of SDE differs radically from that of
conventional stochastic analysis based on martingales.

A fundamental fact on weak convergences in a general setting is as
follows:

\begin{shaded}%
\begin{thm}
\label{thm:WeakApprox}\displabel{thm:WeakApprox}%
{} {\rm (Weak approximation of rough SDE \cite[p.520 Theorem 17.13]{FV10b})}
Assume that 

(i) $\bX_{k}$, $k=1,2,...,\infty$ are $\GEOMR_{[0,T]}^{p}(\R^{d})$-valued
random variables, possibly defined on different probability spaces,
such that $\bX_{k}\to\bX_{\infty}$ $(k\to\infty)$ in law.

(ii) $V\in\Lip^{\gamma}(\R^{e},L(\R^{d},\R^{e}))$, $\gamma>p$, and
$y_{0}\in\R^{e}$. 

Then the $\GEOMR_{[0,T]}^{p}(\R^{d})$-valued random variables $\bY_{k}:=\boldsymbol{\pi}_{(V)}(y_{0},\bX_{k})$
converge to a ${\bf \bY_{\infty}}\in\GEOMR_{[0,T]}^{p}(\R^{d})$ as
$k\to\infty$ in law.\end{thm}
\end{shaded}

\section{Smooth path integral}

\label{sec:pathInt-smooth}\displabel{sec:pathInt{-}smooth}

Assume the compact connected, simply connected semisimple Lie group
$\rG$ is embedded in the matrix Lie group ${\rm GL}(\nu,\C)\subset\Mat(\nu)\cong\C^{\nu^{2}}$.
Then a Brownian motion on $\rG$ is viewed as a process on the Euclidean
space $\C^{\nu^{2}}\cong\R^{2\nu^{2}}$.

Let $\left\{ B(t):\ t\in[0,\infty)\right\} $ be a standard Brownian
motion on $\VECSP=\fg$ w.r.t. the inner product $\left\langle \bullet|\bullet\right\rangle _{\fg}=-\Killing(\bullet,\bullet)$
with $B(0)=0$. Then a Brownian motion $X$ on $\rG$ can be constructed
by the left (resp. right) invariant stochastic differential equation
(SDE)
\begin{equation}
\d X(t)=X(t)\du B(t),\quad\text{resp. }\d X(t)=\left(\du B(t)\right)X(t),\quad X(0)=X_{0}.\label{eq:dXXdW}\displabel{eq:dXXdW}
\end{equation}
where $X_{0}$ is a $\rG$-valued random variable whose distribution
is the Haar measure on $\rG$. Let $\symb{\cV_{\Lft}}\index{${VL}$@${\cV_{\Lft}}$}\hypertarget{ref.125}{}$ (resp.
$\symb{\cV_{\Rght}}\index{${VR}$@${\cV_{\Rght}}$}\hypertarget{ref.126}{}$)$\in\Lip^{\gamma}({\rm Mat}(\nu),L({\rm Mat}(\nu),{\rm Mat}(\nu)))\cap C^{\infty}$,
and assume
\begin{equation}
\cV_{\Lft}(U)M=UM,\quad\cV_{\Rght}(U)M=MU,\quad\forall U\in\rG,\ \forall M\in{\rm Mat}(\nu).\label{eq:def:cV-1}\displabel{eq:def:cV{-}1}
\end{equation}
While $\cV_{\Lft}(A)$ (resp. $\cV_{\Rght}(A)$) is defined for all
$A\in\Mat(\nu)\cong\R^{2\nu^{2}}$, our concern is about the values
on $\rG\subsetneq\Mat(\nu)$ only. Then we can rewrite (\ref{eq:dXXdW})
as a usual SDE on a Euclidean space:
\[
\d X(t)=\cV_{\Lft}(X(t))\du B(t),\text{ resp. }\d X(t)=\cV_{\Rght}(X(t))\du B(t),\quad X(0)=X_{0}\in\rG.
\]
In the following we consider only the left SDE $dX_{t}=\cV_{\Lft}(X_{t})\du B_{t}$.

For a normed vector space $\VECSP$, let
\[
\symb{\GEOMR_{{\rm loc},[0,\infty)}^{p}(\VECSP)}\index{${OmegaGlocp}$@${\GEOMR_{{\rm loc},[0,\infty)}^{p}(\VECSP)}$}\hypertarget{ref.127}{}:=\left\{ {\bf x}:[0,\infty)\to\bbG_{\left\lfloor p\right\rfloor }(\VECSP):\ {\bf x}|_{[0,T]}\in\GEOMR_{[0,T]}^{p}(\VECSP),\ \forall T>0\right\} .
\]
with semi-metrics $\tilde{d}_{p\hvar;[0,T]}$, $T>0$. Note that $\GEOMR_{{\rm loc},[0,\infty)}^{p}(\VECSP)$
is a Polish space. The full It\^o--Lyons map $\bpi_{(\cV_{\Lft})}$
is naturally extended to a map 
\[
\bpi_{(\cV_{\Lft})}:\fg\times\GEOMR_{{\rm loc},[0,\infty)}^{p}\left(\fg\right)\to\GEOMR_{{\rm loc},[0,\infty)}^{p}\left(\Mat(\nu)\right).
\]
Let $\bB$ be the $\GEOMR_{{\rm loc},[0,\infty)}^{p}(\fg)$-valued
random variable which is the enhanced Brownian motion of $B$ on $\fg$,
that is, $\bB^{T}:=\bB|_{[0,T]}\in\GEOMR_{[0,T]}^{p}(\fg)$ is the
enhanced Brownian motion for all $T>0$. Let

\[
\symb{\bX_{0}}\index{${X0}$@${\bX_{0}}$}\hypertarget{ref.128}{}:=1\oplus X_{0}\oplus\left(\frac{1}{2}X_{0}\otimes X_{0}\right)\in\bbG_{2}(\Mat(\nu)).
\]
Then

\[
\bX:=\bpi_{(\cV_{\Lft})}\left(\bX_{0},\bB\right)
\]
is a $\GEOMR_{{\rm loc},[0,\infty)}^{p}(\Mat(\nu))$-valued random
variable, and $\projection_{1}(\bX)$ is identified with the Brownian
motion $X$ on $\rG$. Let $\symb{\mu_{\bX}}\index{${muX}$@${\mu_{\bX}}$}\hypertarget{ref.129}{}$ be the probability
measure on $\GEOMR_{{\rm loc},[0,\infty)}^{p}(\Mat(\nu))$ which is
the law of $\bX$, i.e. $\symb{\mu_{\bX}}\index{${muX}$@${\mu_{\bX}}$}\hypertarget{ref.130}{}:=\bX_{*}\Prob$.

\begin{flushleft}
For $\alpha\in\fg_{\C}^{*}$, let $\symb{V_{\alpha}}\index{${Valpha}$@${V_{\alpha}}$}\hypertarget{ref.131}{}\in\Lip^{\gamma}({\rm Mat}(\nu),L({\rm Mat}(\nu),\C))\cap C^{\infty}$
be such that
\[
V_{\alpha}(g)x=\alpha_{g}^{\Rght}(x),\quad x\in T_{g}\rG,\ g\in\rG
\]
where the tangent space $T_{g}\rG$ is naturally embedded in $\Mat(\nu)$,
and our concern is the values of $V_{\alpha}(g)x$ for $x\in T_{g}\rG\subsetneq\Mat(\nu)$,
$g\in\rG\subsetneq\Mat(\nu)$ only. Then we have
\begin{equation}
\int_{0}^{t}\alpha^{\Rght}(\du X_{s})=\projection_{1}\int_{0}^{t}V_{\alpha}(X_{s})\d\bX_{s}\quad{\rm a.s.}\ \label{eq:intalphaR=00003Dpr}\displabel{eq:intalphaR=00003Dpr}
\end{equation}
where the rhs is a rough line integral.
\par\end{flushleft}

Recall that if $\nu_{n}$, $n=1,...,\infty$ is a sequence of probability
measures on $\GEOMR_{{\rm loc},[0,\infty)}^{p}(\Mat(\nu))$, and if
$\lim_{n\to\infty}\int fd\nu_{n}=\int fd\nu_{\infty}$ for all continuous
and bounded function $f:\GEOMR_{{\rm loc},[0,\infty)}^{p}(\Mat(\nu))\to\R$,
then we say that $\nu_{n}$ weakly converges to $\nu_{\infty}$.

Recall the setting in Sec.\ref{sec:mainTheorem}; $\mu^{1}$ is a
probability measure on $C([0,\infty),\rG)$ which represents a Brownian
motion on $\rG$ (i.e. a Wiener measure uniform on $\rG$). For $h\in C^{\infty}(\rG\cdot\Hwvproj,\R)$
and a (smooth or Brownian) path $\psi:[0,\infty)\to\rG$, let
\[
\symb{\cI_{t}(h;\psi)}\index{${Ithphi}$@${\cI_{t}(h;\psi)}$}\hypertarget{ref.132}{}:=\int_{0}^{t}\alpha^{\Rght}(\du\psi(s))-\im\int_{0}^{t}\fextend h(\psi(s))\d s,\quad\symb{\alpha}\index{${alph}$@${\alpha}$}\hypertarget{ref.133}{}:=-(\lambda+\rho),
\]
where if $\psi$ is smooth, the integral $\int_{0}^{t}\alpha(\du\psi(s))$
is of Riemann--Stieltjes, and if $\psi$ is Brownian, it is a Stratonovich
line integral (or a rough line integral). 

\begin{shaded}%
\begin{prop}
\label{thm:liminttoint}\displabel{thm:liminttoint}Let $\mu_{n}$, $n\in\N$ be probability measures
on $C^{\infty}\left([0,\infty),\rG\right)$$\subset C^{\infty}\left([0,\infty),\Mat(\nu)\right)$.
Define the probability measures $\mu_{n}'$, $n\in\N$ on $\GEOMR_{{\rm loc},[0,\infty)}^{p}(\Mat(\nu))$
by
\begin{equation}
\symb{\mu_{n}'}\index{${mun'}$@${\mu_{n}'}$}\hypertarget{ref.134}{}:=\left(S_{\left\lfloor p\right\rfloor }\right)_{*}\mu_{n},\text{ i.e. }\symb{\mu_{n}'}\index{${mun'}$@${\mu_{n}'}$}\hypertarget{ref.135}{}(E):=\mu_{n}\left(S_{\left\lfloor p\right\rfloor }^{-1}(E)\right).\label{eq:def:mun'}\displabel{eq:def:mun{'}}
\end{equation}
If $\mu_{n}'$ weakly converges to $\mu_{\bX}$, %
{} then we have
\begin{align}
 & \lim_{n\to\infty}\int_{C^{\infty}([0,\infty),\rG)}\xi(\psi)\d\mu_{n}(\psi)=\int_{C([0,\infty),\rG)}\xi(\psi)\d\mu^{1}(\psi),\ \label{eq:liminttoint}\displabel{eq:liminttoint}
\end{align}
where $\symb{\xi(\psi)}\index{${xi(psi)}$@${\xi(\psi)}$}\hypertarget{ref.136}{}:=e^{\cI_{t}(h;\psi)}\ol{\tilde{u}(\psi_{0})}\tilde{v}(\psi_{t}),$
for all $u,v\in V_{\lambda}$ and $t\ge0$. \end{prop}
\end{shaded}
\begin{proof}
Recall that a rough line integral $\int_{0}^{t}\alpha(\d\psi(s))$
is defined by (\ref{eq:intalphaR=00003Dpr}) and (\ref{eq:def:roughInt})
with the full It\^o--Lyons map $\bpi_{(\cV_{\Lft})}$. Since the
$\C$-valued random variable $\xi$ is continuous and bounded for
each $u,v\in V_{\lambda}$, the $n$th law of $\xi$, i.e. $\xi_{*}\mu_{n}$,
weakly converges to $\xi_{*}\mu^{1}$ by Theorem \ref{thm:WeakApprox}.
Hence (\ref{eq:liminttoint}) follows.
\end{proof}
For $\ss>0$, define the probability measure $\symb{\d\mu_{n}^{\ss}(\psi)}\index{${munnu}$@${\d\mu_{n}^{\ss}(\psi)}$}\hypertarget{ref.137}{}$
on $C^{\infty}([0,\infty),\rG)$ to be the time rescaling of $\mu_{n}$
given by $\psi\mapsto\psi_{\ss}:=\psi(\ss^{-1}\bullet)$. Fix an arbitrary
$v_{1}\in V_{\lambda}$ with $\left\Vert v_{1}\right\Vert =1$, and
set

\[
\symb{Z_{\lambda,t,\ss,n}}\index{${Zlambt}$@${Z_{\lambda,t,\ss,n}}$}\hypertarget{ref.138}{}:=\int_{C^{\infty}([0,\infty),\rG)}\left[e^{\cI_{t}(0;\psi)}\ol{\tilde{v}_{1}(\psi_{0})}\tilde{v}_{1}(\psi_{t})\right]\d\mu_{n}^{\ss}(\psi).
\]
By Corollary \ref{thm:main-Brown2} and Proposition \ref{thm:liminttoint},
we find that for any $u,v\in V_{\lambda}$ and $t>0,$ 
\begin{equation}
\bigl\langle u|e^{\im t\cQ(h)}v\bigr\rangle=\lim_{\ss\to\infty}\lim_{n\to\infty}\int_{C^{\infty}([0,\infty),\rG)}\left[e^{\cI_{t}(h)}\ol{\tilde{u}(\psi_{0})}\tilde{v}(\psi_{t})\right]\frac{\d\mu_{n}^{\ss}(\psi)}{Z_{\lambda,t\ss,n}}.\ 
\end{equation}
Thus we find the following theorem:

\begin{shaded}%
\begin{thm}
[Smooth path integral] Let $\mu_{n}$, $n\in\N$ be probability measures
on $C^{\infty}\left([0,\infty),\rG\right)$$\subset C^{\infty}\left([0,\infty),\Mat(\nu)\right)$,
and define $\mu_{n}'$ by (\ref{eq:def:mun'}). If $\mu_{n}'$ weakly
converges to $\mu_{\bX}$ fast enough, then
\begin{equation}
\bigl\langle u|e^{\im t\cQ(h)}v\bigr\rangle=\lim_{n\to\infty}\int_{C^{\infty}([0,\infty),\rG)}\left[e^{\cI_{t}(h)}\ol{\tilde{u}(\psi_{0})}\tilde{v}(\psi_{t})\right]\d\tilde{\mu}_{n}.\ \label{eq:main-smooth2}\displabel{eq:main{-}smooth2}
\end{equation}
where $\tilde{\mu}_{n}$ is the finite measure on $C^{\infty}([0,\infty),\rG)$
given by%
\[
\d\tilde{\mu}_{n}:=\frac{\d\mu_{n}^{n}}{Z_{\lambda,t,n,n}}.
\]
\end{thm}
\end{shaded}

In the above statement, ``$\mu_{n}'$ weakly converges to $\mu_{\bX}$
fast enough'' means precisely that if $\nu_{n}$ $(n\in\N)$ are
probability measures on $C^{\infty}\left([0,\infty),\rG\right)$,
and if $\nu_{n}'$ weakly converges to $\mu_{\bX}$, then there exists
a function $f:\N\to\N$ increasing fast enough such that $\mu_{n}:=\nu_{f(n)}$
satisfy (\ref{eq:main-smooth2}). Thus this condition is neither quantitative
nor constructive; It is an open problem to give a quantitative condition
for (\ref{eq:main-smooth2}).

% ref. http://rexpit.blog29.fc2.com/blog-entry-124.html
%\clearpage
\phantomsection	% ← hyperref を使う時は入れる。使わないなら不要。 	
\addcontentsline{toc}{section}{Reference}

\bibliographystyle{plain}
\bibliography{ybib}

\pagebreak{}

% ref. http://rexpit.blog29.fc2.com/blog-entry-124.html
\clearpage
\phantomsection	% ← hyperref を使う時は入れる。使わないなら不要。 	
\def\indexname{Index (direct href)}
\addcontentsline{toc}{section}{\indexname}
\hypertarget{\indexname}{}
\begin{theindex}

\item $(\alpha,\beta){}$ \href{#ref.3}{*} 
\item $\left\Vert {\bf x}\right\Vert _{p\hvar;[0,T]}{}$ \href{#ref.113}{*} 
\item $A{}$ \href{#ref.80}{*} 
\item $\alpha{}$ \href{#ref.133}{*} \href{#ref.27}{*} \href{#ref.31}{*} \href{#ref.85}{*} 
\item $\alpha^{\Rght}{}$ \href{#ref.24}{*} 
\item $\alpha^{\vee}{}$ \href{#ref.10}{*} 
\item $\frakb^{-}{}$ \href{#ref.33}{*} 
\item $C_{0}^{0,p\hvar}\big([0,T],G^{\left\lfloor p\right\rfloor }\big(\VECSP\big)\big){}$ \href{#ref.115}{*} 
\item $C^{1\hvar}([0,T],\VECSP){}$ \href{#ref.98}{*} 
\item $\Casi{}$ \href{#ref.50}{*} 
\item $\Casi_{+}{}$ \href{#ref.48}{*} 
\item $\Casi_{-}{}$ \href{#ref.47}{*} 
\item $\Casi_{0}{}$ \href{#ref.49}{*} 
\item $C^{p\hvar}([0,T],\cG){}$ \href{#ref.110}{*} 
\item $C_{0}^{p\hvar}([0,T],\cG){}$ \href{#ref.112}{*} 
\item $c_{\lambda}{}$ \href{#ref.86}{*} 
\item $\Delta{}$ \href{#ref.51}{*} 
\item $\dissect[s,t]{}$ \href{#ref.107}{*} 
\item $\Delta^{\alpha}{}$ \href{#ref.57}{*} 
\item $\Delta_{\Rght}^{\alpha}{}$ \href{#ref.58}{*} 
\item $\Delta_{\frakt}^{\alpha}{}$ \href{#ref.60}{*} \href{#ref.61}{*} 
\item $\Delta_{\frakt}{}$ \href{#ref.62}{*} 
\item $\Delta^{\theta}{}$ \href{#ref.54}{*} 
\item $\ul{\Delta}^{-(\lambda+\rho)}{}$ \href{#ref.75}{*} 
\item $d_{\infty;[0,T]}({\bf x},{\bf y}){}$ \href{#ref.122}{*} 
\item $\d^{\alpha}{}$ \href{#ref.55}{*} 
\item $\d^{\theta}{}$ \href{#ref.53}{*} 
\item $d_{{\rm CC}}(g,h){}$ \href{#ref.106}{*} 
\item $\delta_{\lambda,g}{}$ \href{#ref.67}{*} 
\item $\tilde{d}_{\cG,p\hvar;[0,T]}({\bf x},{\bf y}){}$ \href{#ref.111}{*} 
\item $d_{\cG,p\hvar;[0,T]}({\bf x},{\bf y}){}$ \href{#ref.109}{*} 
\item $d_{\lambda}{}$ \href{#ref.20}{*} 
\item $\d\pi_{\lambda}{}$ \href{#ref.13}{*} 
\item $\d_{\Rght}^{\alpha}{}$ \href{#ref.56}{*} 
\item $E_{\alpha}{}$ \href{#ref.35}{*} 
\item $\Hwvproj{}$ \href{#ref.15}{*} 
\item $\Hwvproj(f){}$ \href{#ref.71}{*} 
\item $\Hwvproj(g){}$ \href{#ref.69}{*} 
\item $\eta(t){}$ \href{#ref.79}{*} 
\item $f(\Hwvproj){}$ \href{#ref.70}{*} 
\item $f_{\ss,t}{}$ \href{#ref.77}{*} 
\item $\fextend h{}$ \href{#ref.28}{*} 
\item $\rG\cdot\Hwvproj{}$ \href{#ref.17}{*} 
\item $\hat{\rG}{}$ \href{#ref.1}{*} 
\item $\bbG_{N}(\VECSP){}$ \href{#ref.96}{*} \href{#ref.97}{*} 
\item $\GEOMR_{[0,T],0}^{p}(\VECSP){}$ \href{#ref.114}{*} 
\item $g\cdot\Hwvproj{}$ \href{#ref.68}{*} \href{#ref.16}{*} 
\item $\fg_{\C}^{\alpha}{}$ \href{#ref.4}{*} 
\item $\fg_{N}(\VECSP){}$ \href{#ref.95}{*} 
\item $g_{\ss,t}{}$ \href{#ref.78}{*} 
\item $H(\alpha,V){}$ \href{#ref.82}{*} 
\item $\cH_{\lambda}(\rG){}$ \href{#ref.46}{*} 
\item $\fextend h{}$ \href{#ref.18}{*} \href{#ref.72}{*} 
\item $\bUpsilin_{(\varphi)}({\bf x}){}$ \href{#ref.123}{*} \href{#ref.124}{*} 
\item $\cI_{\alpha,V}{}$ \href{#ref.83}{*} 
\item $\cI_{t}(h){}$ \href{#ref.88}{*} 
\item $\cI_{t}(h;B){}$ \href{#ref.26}{*} 
\item $\cI_{t}(h;\psi){}$ \href{#ref.132}{*} 
\item $\cI_{t}(h;\varphi){}$ \href{#ref.30}{*} 
\item $\Lip^{\gamma}(\R^{d_{1}},L(\R^{d_{2}},\R^{d_{3}})){}$ \href{#ref.118}{*} 
\item $\Lft_{u,v}(g){}$ \href{#ref.64}{*} 
\item $\ell{}$ \href{#ref}{*} 
\item $\mu^{1}{}$ \href{#ref.22}{*} 
\item $\mu^{\ss}{}$ \href{#ref.84}{*} 
\item $\mu_{n}'{}$ \href{#ref.134}{*} \href{#ref.135}{*} 
\item $\d\mu_{n}^{\ss}(\psi){}$ \href{#ref.137}{*} 
\item $\mu^{\ss}{}$ \href{#ref.23}{*} 
\item $\mu_{\bX}{}$ \href{#ref.129}{*} \href{#ref.130}{*} 
\item $\nu{}$ \href{#ref.2}{*} 
\item $\GEOMR_{{\rm loc},[0,\infty)}^{p}(\VECSP){}$ \href{#ref.127}{*} 
\item $\bOmega\bG_{0}^{p}([0,T],\VECSP){}$ \href{#ref.116}{*} 
\item $\GEOMR_{[0,T]}^{p}(\VECSP){}$ \href{#ref.117}{*} 
\item $P{}$ \href{#ref.9}{*} 
\item $P_{\lambda}{}$ \href{#ref.74}{*} 
\item $\phi_{\lambda}{}$ \href{#ref.45}{*} \href{#ref.66}{*} 
\item $\bpi_{(V)}({\bf y}_{0},{\bf x}){}$ \href{#ref.121}{*} 
\item $\pi_{(V)}(y_{0},{\bf x}){}$ \href{#ref.120}{*} 
\item $\pi_{(V)}(y_{0},x){}$ \href{#ref.119}{*} 
\item $\pi_{\lambda}{}$ \href{#ref.14}{*} 
\item $\projection_{\le N}{}$ \href{#ref.94}{*} 
\item $\projection_{k}{}$ \href{#ref.93}{*} 
\item $\cQ(h){}$ \href{#ref.19}{*} \href{#ref.73}{*} 
\item $\Roots{}$ \href{#ref.5}{*} 
\item $\Roots^{+}{}$ \href{#ref.6}{*} 
\item $\Roots_{{\rm s}}^{+}{}$ \href{#ref.8}{*} 
\item $\Roots^{-}{}$ \href{#ref.7}{*} 
\item $\Rght_{u,v}(g){}$ \href{#ref.65}{*} 
\item $\rho{}$ \href{#ref.25}{*} 
\item $S_{N}(x)_{s,t}{}$ \href{#ref.104}{*} 
\item $S_{\ss}(h){}$ \href{#ref.87}{*} 
\item $\spec{}$ \href{#ref.63}{*} 
\item $\Tensor(\VECSP){}$ \href{#ref.90}{*} 
\item $T_{1},...,T_{\ell}{}$ \href{#ref.34}{*} 
\item $\Tensor^{k}(\VECSP){}$ \href{#ref.91}{*} 
\item $\Ltrans{}$ \href{#ref.36}{*} 
\item $\Tensor^{\le N}(\VECSP){}$ \href{#ref.92}{*} 
\item $T_{\ss}{}$ \href{#ref.76}{*} 
\item $\Rtrans{}$ \href{#ref.37}{*} 
\item $T_{\ss}(V){}$ \href{#ref.81}{*} 
\item $\scU(\fg_{\C}){}$ \href{#ref.44}{*} 
\item $\frakn^{-}{}$ \href{#ref.32}{*} 
\item $V_{\alpha}{}$ \href{#ref.131}{*} 
\item $\cV_{\Lft}{}$ \href{#ref.125}{*} 
\item $\cV_{\Rght}{}$ \href{#ref.126}{*} 
\item $\tilde{v}{}$ \href{#ref.21}{*} 
\item $\cX(\rT){}$ \href{#ref.11}{*} 
\item $\cX_{+}(\rT){}$ \href{#ref.12}{*} 
\item $\bX_{0}{}$ \href{#ref.128}{*} 
\item $X_{k}{}$ \href{#ref.52}{*} 
\item $X^{\Rght}{}$ \href{#ref.39}{*} 
\item $X^{\Lft}{}$ \href{#ref.38}{*} \href{#ref.40}{*} 
\item $X^{\Rght}{}$ \href{#ref.41}{*} 
\item $x_{s,t}^{\{\le N\}}{}$ \href{#ref.105}{*} 
\item $x_{s,t}^{\{0\}}{}$ \href{#ref.99}{*} 
\item $x_{s,t}^{\{1\}}{}$ \href{#ref.101}{*} 
\item $x_{s,t}^{\{2\}}{}$ \href{#ref.103}{*} 
\item $x_{s,t}^{\{n\}}{}$ \href{#ref.100}{*} 
\item $\xi(\psi){}$ \href{#ref.136}{*} 
\item $\xi_{k}{}$ \href{#ref.59}{*} 
\item ${\bf x}_{s,t}{}$ \href{#ref.108}{*} 
\item $x_{s,t}{}$ \href{#ref.102}{*} 
\item $Z_{\lambda,t,\ss}{}$ \href{#ref.29}{*} \href{#ref.89}{*} 
\item $Z_{\lambda,t,\ss,n}{}$ \href{#ref.138}{*} 
\item $Z^{\Lft}{}$ \href{#ref.42}{*} 
\item $Z^{\Rght}{}$ \href{#ref.43}{*} 
\item $\frakt$-partial magnetic Laplacian \href{#_24_5Cfrakt_24-partial+magnetic+Laplacian}{*} 
\item $p$-absolutely continuous \href{#_24p_24-absolutely+continuous}{*} 
\item $p$-variation distance \href{#_24p_24-variation+distance}{*} 
\item $p$-variation metric \href{#_24p_24-variation+metric}{*} 
\item absolutely continuous of order $p$ \href{#absolutely+continuous+of+order+_24p_24}{*} 
\item adjoint operation \href{#adjoint+operation}{*} 
\item algebraically integral weight \href{#algebraically+integral+weight}{*} 
\item analytically integral weight \href{#analytically+integral+weight}{*} 
\item Borel--Weil theorem \href{#Borel--Weil+theorem}{*} 
\item Carnot--Carath\'eodory metric \href{#Carnot--Carath_5C_27eodory+metric}{*} 
\item Casimir element \href{#Casimir+element}{*} 
\item character lattice \href{#character+lattice}{*} 
\item coherent states \href{#coherent+states}{*} 
\item coroot \href{#coroot}{*} 
\item delta function \href{#delta+function}{*} 
\item dominant weights \href{#dominant+weights}{*} 
\item enhanced Brownian motion \href{#enhanced+Brownian+motion}{*} 
\item finite $p$-variation \href{#finite+_24p_24-variation}{*} 
\item free nilpotent group \href{#free+nilpotent+group}{*} 
\item full It\^o--Lyons map \href{#full+It_5C_5Eo--Lyons+map}{*} 
\item full RDE solution \href{#full+RDE+solution}{*} 
\item Glauber--Sudarshan-type quantization \href{#Glauber--Sudarshan-type+quantization}{*} \href{#Glauber--Sudarshan-type+quantization.1}{*} 
\item GS quantization \href{#GS+quantization}{*} \href{#GS+quantization.1}{*} 
\item geometric $p$-rough path \href{#geometric+_24p_24-rough+path}{*} 
\item horizontal \href{#horizontal}{*} 
\item It\^o--Lyons map \href{#It_5C_5Eo--Lyons+map}{*} 
\item Laplacian \href{#Laplacian}{*} 
\item magnetic exterior differentiation \href{#magnetic+exterior+differentiation}{*} 
\item magnetic Laplacian \href{#magnetic+Laplacian}{*} 
\item orbit \href{#orbit}{*} 
\item positive root \href{#positive+root}{*} 
\item pre-Borel--Weil theorem \href{#pre-Borel--Weil+theorem}{*} 
\item rank \href{#rank}{*} 
\item reproducing kernel \href{#reproducing+kernel}{*} 
\item roots \href{#roots}{*} 
\item rough line integral \href{#rough+line+integral}{*} 
\item simple roots \href{#simple+roots}{*} 
\item solution of the rough differential equation (RDE solution) \href{#solution+of+the+rough+differential+equation+_28RDE+solution_29}{*} 
\item step-$N$ signature \href{#step-_24N_24+signature}{*} 
\item truncated tensor algebra \href{#truncated+tensor+algebra}{*} 
\item variation seminorm \href{#variation+seminorm}{*} 
\item Weyl's canonical basis \href{#Weyl_27s+canonical+basis}{*} 
\item weak geometric $p$-rough paths \href{#weak+geometric+_24p_24-rough+paths}{*} 
\item weight lattice \href{#weight+lattice}{*} 
\end{theindex}

\end{document}